%% file: gen_mp.tex
\documentclass{article}
\usepackage{ijcai26}

\usepackage{times}
\usepackage{soul}
\usepackage{xurl}
\usepackage[hidelinks]{hyperref}
\usepackage[utf8]{inputenc}
\usepackage[small]{caption}
\usepackage{graphicx}
\usepackage{amsmath, amsthm, amssymb}
\usepackage{booktabs}
\usepackage{algorithm}
\usepackage{algorithmic}
\usepackage[switch]{lineno}

\usepackage{algorithm}
\usepackage{algorithmic}

\usepackage{listings}
\makeatletter
\newcommand{\alglinelabel}[1]{%
  \protected@write\@auxout{}{%
    \string\newlabel{#1}{{\arabic{ALC@line}}{\thepage}}%
  }%
}
\makeatother

\usepackage{pifont} 

\newcommand{\cmark}{\ding{51}}

\usepackage{tikz}
\definecolor{lightblue}{rgb}{0.68, 0.85, 0.9}
\tikzstyle{greybox} = [draw=black, fill=lightblue!20, very thick,
    rectangle, rounded corners, inner sep=10pt, inner ysep=15pt]
\tikzstyle{fancytitle} =[fill=black, text=white, rounded corners]    

\title{Grammar-Aware Literate Generative Mathematical Programming with Compiler-in-the-Loop}

\author{
Roberto Rossi$^1$
\and
Steven D. Prestwich$^2$
\affiliations
$^1$Business School, University of Edinburgh, UK\\
$^2$Insight Centre for Data Analytics, University College Cork, Cork, Ireland
\emails
roberto.rossi@ed.ac.uk,
s.prestwich@ucc.ie
}

\begin{document}

\maketitle

\begin{abstract}
Mathematical programming is widely employed across various sectors --- such as logistics, energy, and workforce planning --- to model and solve industrial optimisation problems, but its use requires substantial domain expertise. Large language models offer a promising way to translate natural-language problem descriptions into optimisation models, yet existing approaches are costly and generally produce models written in general-purpose computer code (e.g. Python), which can be difficult to inspect, validate, and reuse. 
In this work, we introduce SyntAGM, a system that generates optimisation models in a readable algebraic modelling language through an iterative generate--compile--assess--revise loop. 
SyntAGM leverages PyOPL, an OPL-like modelling language compiler designed to provide actionable feedback for iterative model repair.
To obtain a valid PyOPL model that matches the problem description, SyntAGM mobilises compiler feedback and an LLM-based alignment judge.
In addition, it combines in-context exposure to the target language grammar, and few-shot retrieval of modelling exemplars. 
 Across multiple benchmarks, SyntAGM achieves a more favourable cost--quality trade-off compared to established prompting baselines. 
\end{abstract}

\section{Introduction}

Operational Research (OR) is the discipline of applying advanced analytical methods, such as mathematical programming and simulation, to solve complex organisational problems in a variety of sectors, including supply chain management \cite{Peterson1998-al}, airport operations \cite{Jacquillat2015-ai}, and workforce planning \cite{Jaillet2022-jg}. 

Mathematical programming is a key tool in OR used to formulate real-world problems as optimisation models --- also called mathematical programmes --- that are built on decision variables, constraints, and objective function(s), and that can be solved by using off-the-shelf solvers (e.g. Gurobi \cite{gurobi}). 

\begin{figure}[h!]
\centering
\includegraphics[width=\linewidth]{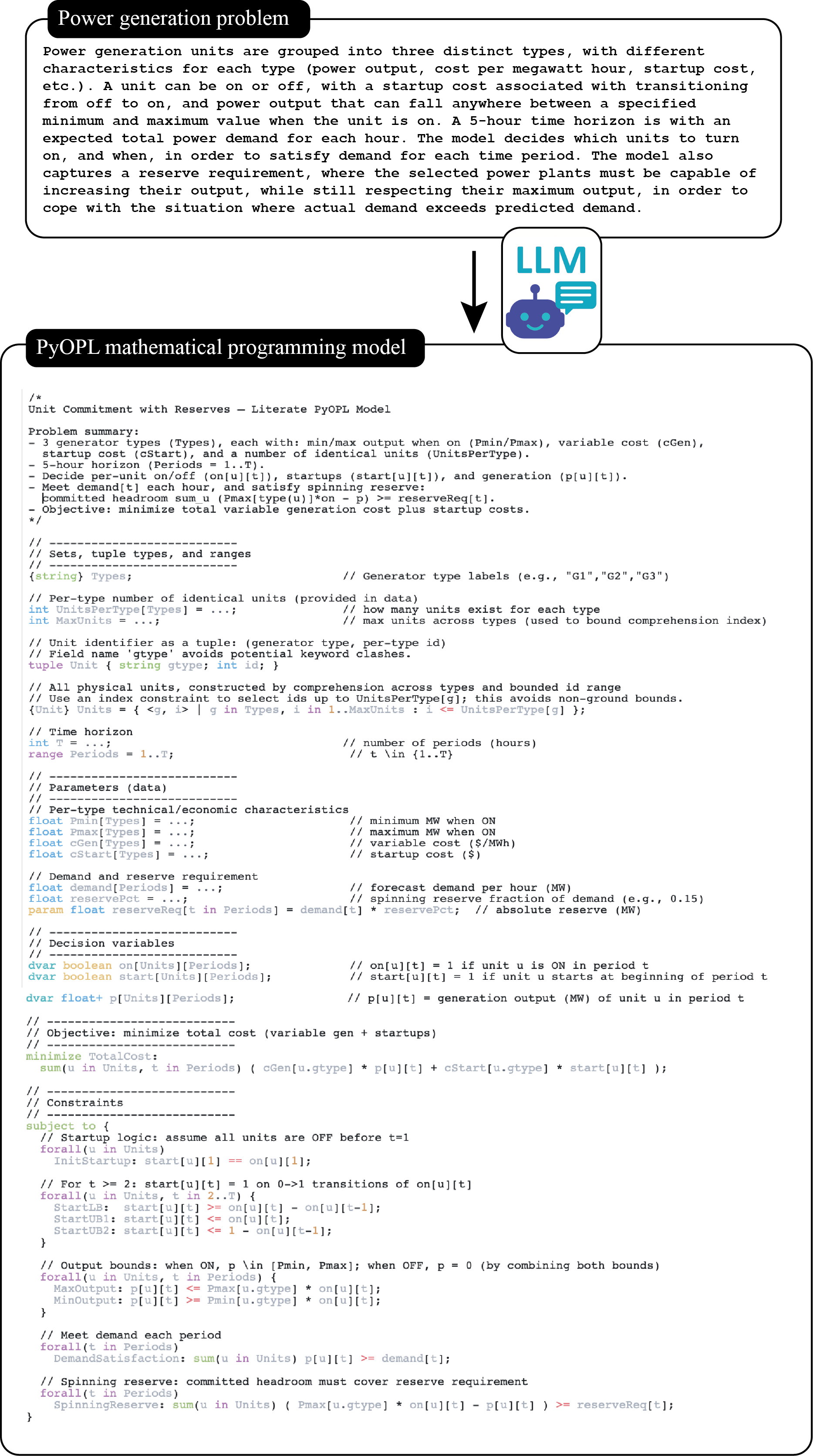}
\caption{A GenMP task --- translating a problem description in the realm of power generation into a mathematical programme.}
\label{fig:genmp_2}
\end{figure}

The application of Large Language Models (LLMs) in OR is a thriving area of research \cite{Fan2025-ny}. Within this domain of research, Generative Mathematical Programming (GenMP) --- which is concerned with translating the natural language description of an optimisation problem into a mathematical programme --- has received increased attention in recent years. This task is challenging because it requires domain-specific knowledge to understand terminology, and because descriptions often contain implicit constraints that must be inferred from context. In Figure \ref{fig:genmp_2} we illustrate the GenMP task of translating a natural-language problem description into a mathematical programme expressed in an Algebraic Modelling Language.

In a recent survey \cite{DBLP:conf/ijcai/XiaoXXGZHFYWSKD25}, the authors review emerging techniques, and highlight opportunities, limitations, and future research directions at the intersection of LLMs and mathematical programming. In their taxonomy, they classify works in three areas: development of evaluation frameworks \cite{xing-etal-2024-towards} and benchmark datasets such as NL4OPT \cite{pmlr-v220-ramamonjison23a} and IndustryOR \cite{Huang2025}; development of domain-specific fine-tuned LLMs \cite{DBLP:conf/iclr/JiangSQLZZY25}; and development of {\em advanced inference frameworks}, including multi-agent systems such as Chain-of-Experts (CoE) \cite{xiao2024chainofexperts}, and Chain-of-Thought (CoT)  \cite{10.5555/3600270.3602070} variants \cite{10.5555/3666122.3666639,10.1609/aaai.v38i16.29720} --- our work sits within this last area.

During {\em inference}, LLMs translate a problem description expressed in natural language into a mathematical programme.
Prompt engineering is a simple and yet effective method that can be operationalised in this context.  
Beyond so-called ``standard prompting,'' in which the LLM is directly prompted to emit the final model, existing studies enhanced LLM capabilities by either encouraging LLMs to generate additional intermediate reasoning steps --- an approach referred to as ``X-of-thought;''  or by developing LLM-based multi-agent systems in which multiple ``expert'' LLMs cooperate to tackle the problem, as seen in CoE and OptiMUS \cite{pmlr-v235-ahmaditeshnizi24a}. Despite the rapid pace of expansion of the literature on advanced inference frameworks, some noticeable gaps remain, which we next discuss.

Algebraic Modelling Languages (AMLs) let users express mathematical programmes in a declarative style that mirrors algebraic notation \cite{Kallrath2004-jy};
examples of AMLs include AMPL, GAMS, and OPL.
AMLs separate model from data and compile to multiple solver back ends, aiding readability, auditability, and solver independence.

{\bf Gap 1.}
{\em Rather than leveraging AMLs, most prior work on GenMP focuses on emitting solver API code} (e.g., \texttt{gurobipy}, \texttt{pulp}), which interleaves programming logic and constraints. 
In practice, requirement elicitation is iterative and human-centred: practitioners interview customers, refine requirements, prototype, solicit feedback, and iterate. 
GenMP can shorten these cycles and support near real-time construction and execution of model components during requirement analysis \cite{Freuder2017,Freuder2024-xc}; but in similar human-in-the-loop settings, outputs must be readable and auditable, and use of AMLs is therefore advantageous. 
Additionally, AML models can be checked statically via compilation; compiler syntax/semantic diagnostics --- particularly when tailored to be expressive and informative --- can drive principled, iterative refinement in GenMP. 

{\bf Gap 2.} 
{\em Prior work does not consider new AMLs whose grammar is unknown to LLMs; nor does it investigate the use of in-context AML grammar paired with compiler diagnostics to tackle this scenario}. 

Most importantly, no study has investigated if detailed and actionable compiler diagnostics are beneficial for GenMP systems.
In fact, we believe that insufficient attention has been devoted to developing effective compiler diagnostics that, rather than simply pointing out that a syntax error exists, explain what causes the error, and how an error can be fixed. This would be of course beneficial to human developers; but in GenMP settings, actionable diagnostics can be immediately turned into refinement prompt, and hence are highly desirable.

{\bf Gap 3.}
{\em Beyond accuracy, there is limited evidence on the trade-off between solution quality, inference cost} (tokens, dollars), {\em and latency} (end-to-end wall-clock timer) {\em for existing methods}. Few studies report telemetry or characterise cost--accuracy trade-offs across prompting strategies. In GenMP, these trade-offs are important given frequent multi-step reasoning, compilation retries, and alignment checks.

{\bf Contributions.}
We make the following contributions to the literature on GenMP.

\begin{itemize}
\item We introduce PyOPL, a Python library for parsing OPL-like models. PyOPL focuses on the core algebra of mathematical programming in OPL \cite{Van_Hentenryck1999-zt} and features {\em actionable error messages} an LLM can leverage to revise a model that fails to compile. 
\item We introduce Syntax-Aware Generative Modelling (SyntAGM), a system that translates natural-language problem descriptions into PyOPL models via an iterative generate--compile--assess--revise workflow. 
The system comprises: (i) a concrete, in-context PyOPL grammar/semantics reference; (ii) top-$k$ RAG to retrieve few-shot exemplars; (iii) \emph{literate}\footnote{{\bf Literate programming} \cite{Knuth1984-at} treats programs --- including mathematical programs --- as explanations for readers by interleaving prose and source code so that the author's modelling rationale is presented as a coherent narrative.} modelling; (iv) a bespoke PyOPL compiler that provides {\em actionable} feedback; (v) an LLM-as-a-Judge alignment assessor that decides whether the model matches the prompt intent; and (vi) a revision prompt that applies minimal edits guided by compiler diagnostics or assessor feedback. 
\item We instrument token usage, dollar cost, runtime, and iteration counts to study the cost--quality frontier of GenMP systems. Whilst achieving comparable quality (i.e. accuracy), SyntAGM offers a more favorable cost--quality trade-off than existing baselines (Standard, Chain-of-Thought, Tree-of-Thoughts, Reflexion, and Chain-of-Experts) on multiple datasets: NL4OPT, ComplexOR, ReSocratic, and IndustryOR. 
\end{itemize}

\section{Related works}

Early attempts to translate natural-language optimisation problems into formal mathematical programs include \cite{ramamonjison-etal-2022-augmenting}, which proposed a two-stage pipeline wherein a BART-based model \cite{lewis-etal-2020-bart} generates an intermediate representation (IR) that is subsequently parsed to a canonical optimisation model.
This line of work also introduced NL4Opt \cite{pmlr-v220-ramamonjison23a}, a benchmark of 1101 annotated Linear Programming Word Problems (LPWPs), and experimented with systematic prompt template design (problem text, task instructions, and format control) in the associated competition.
Following this first attempt, many studies have followed investigating applications of LLMs to GenMP; these studies can be broadly classified in training-based and prompt-based approaches. 

{\bf Training-based methods} leverage data synthesis and instruction tuning to fine-tune open-source LLMs --- such as Mistral and LLaMA --- for optimisation modelling; examples include ORLM \cite{Huang2025}, which proposed a semi-automated process for creating synthetic training data during fine-tuning to address the lack of suitable large-scale training datasets; and LLMOPT \cite{DBLP:conf/iclr/JiangSQLZZY25}, which trains the LLMs to produce a five-element formulation comprising sets, parameters, variables, objective and constraints.  These methods can yield strong adherence to task structure but require substantial data engineering, risk domain drift, and generally target solver APIs rather than AMLs, limiting readability and solver independence.

{\bf Prompt-based methods} utilise existing LLM interfaces to automate solving optimisation problems. Building on the seminal work of \cite{NEURIPS2020_1457c0d6} that pioneered in-context learning, these methods aim to instil domain knowledge into LLMs through the prompt. Following \cite{DBLP:conf/ijcai/XiaoXXGZHFYWSKD25}, prompt-based methods can be further decomposed into two classes: ``X-of-thought'' and ``Multi-expert.''  

{\em X-of-thought} methods encourage LLMs to generate additional intermediate reasoning steps. This approach was pioneered by \cite{10.5555/3600270.3602070}, who introduced the Chain-of-Thought approach, which encourages the LLM to think step-by-step to prevent logical gaps during inference. 
Follow up works include Tree-of-Thoughts \cite{10.5555/3666122.3666639} and Graph-of-Thoughts \cite{10.1609/aaai.v38i16.29720} methods, which employ tree- and graph- structured exploration of reasoning paths. 
Reflexion \cite{NEURIPS2023_1b44b878} leverages linguistic feedback to induce better decision-making: reflexion agents verbally reflect on task feedback signals, and maintain reflective text in an episodic memory buffer. 

{\em Multi-expert} methods aim to scale language models for complex reasoning via multi-agent collaboration systems \cite{DBLP:conf/iclr/QianXW0ZXDDC00025} in which LLMs play the role of experts. OptiMUS \cite{pmlr-v235-ahmaditeshnizi24a} uses a pre-defined workflow to engage experts. Chain-of-Experts \cite{xiao2024chainofexperts} leverages a ``Conductor'' to orchestrate the engagement process, and a system-level reflection mechanism aligned with \cite{NEURIPS2023_1b44b878}. 
These frameworks can tackle complex instances but incur heavy token usage and orchestration overhead.

While proposing new approaches authors also developed {\bf benchmarks} to assess them. In addition to the NL4Opt benchmark, another early benchmark is NLP4LP \cite{pmlr-v235-ahmaditeshnizi24a}. Most problems in these early benchmarks featured low complexity; to address this gap, authors introduced the ComplexOR \cite{xiao2024chainofexperts} and IndustryOR \cite{Huang2025} benchmarks, which cover more complex instances collected from both industrial and academic scenarios. Unfortunately, all these benchmarks originally suffered from quality control issues. ReSocratic \cite{DBLP:conf/iclr/Yang0HGSHFSL025} investigated the use of multiple filters to remove erroneous cases. \cite{DBLP:conf/ijcai/XiaoXXGZHFYWSKD25} manually filtered all error cases from mainstream benchmarks and compiled a unified, cleaned collection of optimisation modelling benchmarks to facilitate future research.

Most existing works on GenMP employ objective-wise {\bf evaluation}, in which the focus is solely on the correctness of the final objective function value obtained by solving the model generated, e.g. \cite{deng2024cafa,xiao2024chainofexperts,Huang2025}. Of course, a correct objective does not guarantee a correct model, therefore alternative approaches have been considered. \cite{pmlr-v220-ramamonjison23a} extract coefficients of the objective function and constraints and compare them with a canonical ground truth representation; unfortunately, because the LLM generates arbitrary variable names and index dimensions, automating the semantic alignment of variables is a non-trivial, open problem in general \cite{zhai2025equivamapleveragingllmsautomatic}. Additionally, similar approaches fail to capture the degree of correctness of a solution; to overcome this limitation graph-based evaluation approaches have been recently investigated \cite{xing-etal-2024-towards}. Despite these developments, as remarked in \cite{DBLP:conf/ijcai/XiaoXXGZHFYWSKD25}, objective-wise evaluation paired with accuracy remains the most widely accepted measure. This strategy is a pragmatic compromise, as it functions as a ``semantic hash'' that bypasses the need for manual variable mapping. 

Finally, cost/latency are still under-reported in the literature, despite their importance for multi-step GenMP pipelines.

\section{Preliminaries}

A mathematical programme is defined as follows
\begin{align*}
  \min_{x} \quad & f(x) \\
  \text{subject to} \quad & g_i(x) \leq 0 \quad \forall i \in \mathcal{I} \\
  & h_j(x) = 0 \quad \forall j \in \mathcal{J}
\end{align*}
where $x \in \mathbb{R}^n$ is the vector of decision variables, $f(x)$ is the objective function, $g_i(x)$ are the inequality constraints, and $h_j(x)$ are the equality constraints. 

In what follows, we focus on the {\em challenge of translating the natural language description of an optimisation problem into a mathematical programme expressed via an AML.} 

There are three key aspects to consider when building a GenMP system that targets an AML:
(i) ensuring syntactic and semantic correctness;
(ii) LLM awareness of AML grammar constructs; and 
(iii) the use of appropriate mathematical modelling patterns. 

{\bf PyOPL.}
A popular AML is the Optimisation Programming Language (OPL), which was introduced over 25 years ago \cite{Van_Hentenryck1999-zt} to simplify the formulation and solution of optimisation problems. 
To support this study, we developed PyOPL, a Python library for parsing OPL-like models. PyOPL focuses on the core algebra of mathematical programming in OPL --- typed declarations, indexed structures, tuple types, sums/forall, basic functions and aggregates --- and adds syntax/semantic checks (e.g. typed set validation, index typing, array shape checks) so that models either compile cleanly to \texttt{gurobipy} \cite{gurobi} or \texttt{scipy} (HiGHS) \cite{Huangfu2018-jc}, or produce {\em actionable diagnostics}. Error messages produced by the PyOPL compiler are specific and informative, this means the LLMs can take full advantage of them while revising a model that fails to compile. 
There are over 170 different types of compilation errors implemented in PyOPL, each providing a specific message describing the nature of the issue that led to the error and {\em how the error can be resolved}. In Table \ref{tab:industryor_errors}, we provide examples of compiler error messages observed while solving problems from the ChallengeOR benchmark.
\begin{table}[htbp]
\centering
\begin{tabular}{lp{0.78\columnwidth}}
\hline
\textbf{ID} & \textbf{Error message} \\
\hline
E1 & Semantic Error (Line 37): Range bounds must be integer. \\
E2 & Semantic Error (Line 19): Syntax error in .dat file at or near token [, value '['. Hint: a common cause is a nested keyed-array block (for example, 'A = [ "k1" [ "sub1" [...], "sub2" [...] ], ... ];'). This parser supports one keyed-array level only: after a string or tuple key, the value must be a scalar or a plain array, not another keyed sub-block. \\
E3 & Semantic Error: Range 'T' was supplied in the data file, but ranges used for indexing must be declared with explicit bounds in the model file. Declare it in the model (e.g., 'range T = 1..N;') and remove it from the .dat. \\
\hline
\end{tabular}
\caption{Sample PyOPL compiler diagnostics observed in ChallengeOR.}
\label{tab:industryor_errors}
\end{table}
PyOPL source code and a table mapping all syntax errors emitted by PyOPL can be found in our supplementary material (SM).

{\bf In-context learning of AML grammar.} Albeit grounded on OPL, PyOPL is a brand new language: it only implements a subset of the OPL language, so one cannot reasonably expect an LLM to know what constructs can or cannot be used in a model. In contrast to other competing GenMP approaches, we do not strictly enforce syntax at decode time via Backus--Naur form (BNF) grammar-constrained decoding \cite{geng-etal-2023-grammar}. 
Conversely, we make the LLM aware of the implemented PyOPL grammar by leveraging in-context learning \cite{NEURIPS2020_1457c0d6,wang2023grammar}. 
In-context learning is the LLM ability to use prompt instructions and examples as implicit training data to infer the task and produce outputs without updating its parameters. 
To this end, we leverage a reference Markdown document --- provided in our SM --- that specifies the concrete BNF grammar and semantics for the PyOPL modelling language. The grammar is annotated with additional plain-text clarifications and relevant semantics, to capture aspects of the language that cannot be fully expressed via the BNF grammar.

{\bf Modelling patterns in MILP.}
{\em Design patterns} are typical solutions to common problems found in software design. Like software development, mathematical modelling is also rich with modelling patterns. While there exists no book comparable to \cite{Gamma1994-mc} in the realm of mathematical modelling, mainstream OR textbooks discuss a wealth of modelling patterns that practitioners use to structure mathematical programs.
 
In Mixed-integer linear programming (MILP), a modelling pattern is a small, reusable formulation template—typically a set of decision variables and linear constraints, sometimes with an objective fragment—that encodes a logical or combinatorial relation (e.g., exactly-one choice, coverage, implication, precedence) which recurs across MILP formulations.

\input{patterns_table.tex}

Informed by the academic literature on mathematical programming, we have identified a range of 25 MILP modelling patterns commonly found in mathematical programs --- these are discussed in details in our Technical Appendix (TA)-I. Given a set of modelling patterns, one may devise a set of problems that embed one or more of these patterns (Table \ref{tab:patterns_by_problem}). 
Synthetic literate implementations of 22 exemplar problems --- each of which comprises description, model, and data --- represents a knowledge base (provided in our SM) against which we perform top-$k$ retrieval to support in-context learning; in addition, this knowledge base can expand as new problem descriptions and associated models become available.

\section{The architecture of SyntAGM}\label{sec:syntagm_architecture}

We cast SyntAGM as a sequential decision process whose state is a prompt context (problem text, PyOPL grammar, few-shot exemplars, and the latest attempt), whose action is to emit a candidate PyOPL model--data pair, and whose environment returns programmatic signals via compilation and alignment checks. 
This perspective clarifies the analogy with actor--evaluator--reflection systems \cite{NEURIPS2023_1b44b878} while highlighting key differences specific to AML-targeted synthesis and compiler-in-the-loop modelling. The complete implementation of SyntAGM is provided in our SM. 

The architecture of SyntAGM comprises four modules.

{\bf Generator $M_g$.}
Given a context $s_t$, the generator --- i.e. the LLM --- samples a candidate $(\texttt{model}, \texttt{data}) \sim \pi_\theta(\cdot \mid s_t)$, where $\pi_\theta$ is the stochastic policy induced by the LLM with parameters $\theta$ over the set of allowed artefacts. 
The generator prompt asks the LLM to do its reasoning in a private CoT-style scratchpad (i.e. ``think step by step'') and to produce the final result --- that is the model--data pair --- in a literate PyOPL style; note that, unlike classic CoT, which outputs its reasoning, the system only returns a JSON object in line with a prescribed output contract (described in our TA-II).
The context further includes a PyOPL Markdown grammar reference and top-$k$ few-shot model exemplars retrieved with RAG from the knowledge base (see TA-III for RAG implementation details) --- note that existing few-shot/RAG works generally inject code-only snippets, not description--model--data triplets.

{\bf Evaluator $M_e$ (compiler- and assessor-driven).}
The evaluator in SyntAGM comprises two elements:
(i) the deterministic PyOPL compiler, and 
(ii) an LLM-based judge that assesses alignment to prompt intent. 
Formally, let
\begin{align*}
&(\,b_t,\,E_t\,) \leftarrow \texttt{compile}(\texttt{model}_t,\texttt{data}_t),\\
&(\,a_t,\,S_t\,) \leftarrow \texttt{assess}(\texttt{model}_t,\texttt{data}_t,\texttt{prompt})
\end{align*}
where $b_t\in\{0,1\}$ denotes successful compilation and $E_t$ is a (possibly empty) set of error messages; 
$a_t\in\{0,1\}$ is a binary flag that signals alignment and $S_t$ is a textual assessment discussing misalignment.
The following binary signal, representing successful compilation and alignment to prompt intent
\[
r_t \;=\; \mathbb{I}\{\,b_t=1 \wedge a_t=1\,\},
\]
is used to determine when to terminate and return the final model--data pair. 
The evaluator thus supplies both a {\em binary gate} and textual feedback to guide revision for both AML syntax errors and prompt intent.

{\bf Self-reflection $M_{sr}$.}
When syntax errors or a misalignment are encountered (i.e. $r_t=0$), the system generates a revision prompt that contains the latest attempt and the evaluator's feedback. 
The prompt either comprises syntax-repair guidance if $E_t\neq\emptyset$, or alignment-repair guidance if $E_t=\emptyset$ but $a_t=0$;
in both cases, it instructs the LLM to implement minimal edits and preserve modelling intent. 
Note that the same LLM backend acts as a revision policy:
\[
(\texttt{model}_{t+1},\texttt{data}_{t+1}) \sim M_{sr}(\cdot \mid s_t,\,E_t,\,S_t).
\]
In practice, $M_{sr}$ and $M_g$ are the same LLM, which is conditioned on different instructions and evidence. 
More specifically, compiler diagnostics and assessor critiques are the pieces of information that direct the next attempt.

{\bf Memory $\mathrm{mem}$.}
The state $s_t$ is the context, which aggregates:
(i) the problem description; 
(ii) the PyOPL grammar reference $G$; 
(iii) a small set $F_k$ of few-shot exemplars (description--model--data) retrieved via RAG; and 
(iv) the most recent attempt plus its diagnostics $(\texttt{model}_t,\texttt{data}_t,E_t,S_t)$.
SyntAGM memory thus lives in prompt sections and is refreshed at each iteration. 
Most importantly, the emitted literate PyOPL artefacts --- the model and data files with labelled objectives/constraints and localised comments --- serve as the agent's {\em long-term} memory. Each round of feedback $(E_t,S_t)$ is summarised in the model and data files via literate comments during revision; rationale is thus ``woven'' --- to use Knuth's own metaphor --- into the code (i.e. the purpose of a parameter/variable, why a domain is chosen, why a constraint is repaired) exactly where it matters. This information not only persists across iterations, but it is also human-auditable. Compared to a separate reflective buffer, such as that implemented in Reflexion \cite{NEURIPS2023_1b44b878}, embedding memory inside the artefacts has two advantages: comments appear next to the precise declarations and constraints they explain; and the resulting model and data files are reusable exemplars that can be added to SyntAGM knowledge base to tackle future problems. 

\begin{figure*}[htbp]
\centering
\includegraphics[width=\linewidth]{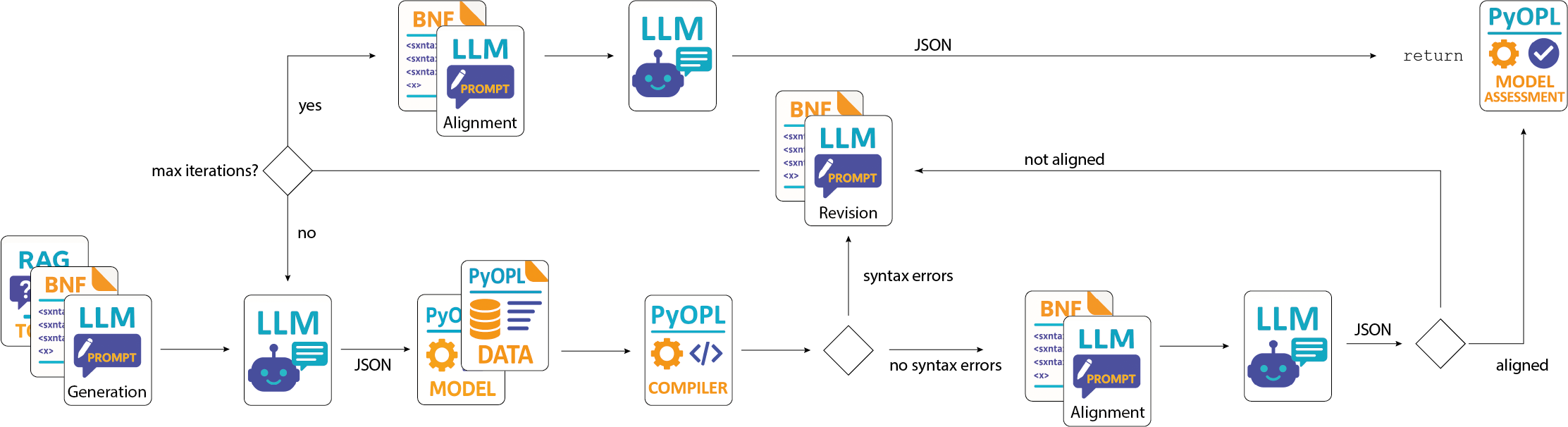}
\caption{High-level generate–compile–assess–revise loop of SyntAGM.}
\label{fig:syntagm}
\end{figure*}

{\bf The SyntAGM loop.}
Putting the components together, SyntAGM executes the loop (generate--compile--assess--revise) illustrated in Figure~\ref{fig:syntagm}, which comprises the following four steps.

\begin{enumerate}
  \item Initialize $s_0 \leftarrow (\text{problem}, G, F_k)$; query $M_g$ to obtain $(\texttt{model}_0,\texttt{data}_0)$.
  \item Evaluate: $(b_t,E_t)\leftarrow \texttt{compile}(\texttt{model}_t,\texttt{data}_t)$.
        If $b_t=1$, run alignment $(a_t,S_t)\leftarrow \texttt{assess}(\cdot)$.
  \item If $r_t=\mathbb{I}\{b_t=1 \wedge a_t=1\}=1$, stop and return the aligned artefacts and assessment.
  \item Else, build a revision prompt from $(E_t,S_t)$ and obtain $(\texttt{model}_{t+1},\texttt{data}_{t+1}) \sim M_{sr}(\cdot)$; update $s_{t+1}$ and repeat until the iteration budget is reached.
\end{enumerate}
In compact notation,
{\tiny
\[
(\texttt{model}_{t+1},\texttt{data}_{t+1})
\;\sim\;
\pi_\theta\!\left(\,\cdot \;\middle|\; \underbrace{\text{problem}}_{\text{task}},
\underbrace{G}_{\text{syntax/semantics}}, 
\underbrace{F_k}_{\text{structural priors}},
\underbrace{E_t,S_t}_{\text{reflective feedback}} \right),
\]
}
with termination on the first $t$ such that $r_t=1$.

Implementation details, pseudocode of SyntAGM, and a worked example are provided in TA-IV and TA-V, respectively.

{\bf Analogies and differences to Reflexion-style agents.}
SyntAGM proceeds by trial-and-error: the LLM crafts a solution (generate), an evaluator gates progress and provides textual feedback (evaluate), and lessons from one attempt to another are carried forward in the prompt state (revise); in this respect, it is similar to Reflexion-style agents \cite{NEURIPS2023_1b44b878}. However, SyntAGM also benefits from deterministic compiler output messages that can be leveraged to address errors, and from the assessment provided by an LLM judge, which guides convergence towards the final goal --- SyntAGM loop is {\em curricular} \cite{bengio_curriculum_2009}: make sure that the code compiles, and only then check alignment.
In addition, in SyntAGM the prompt features a concrete AML grammar and a strict JSON output contract. 
``Memory'' accumulates inside the PyOPL artefacts thanks to the literate comments, rather than in a separate reflective buffer. 
 This architecture prioritises AML correctness, auditability, and cost control over an open-ended debate among LLM agents --- as observed, for instance, in CoE.

{\bf Prompt templates.}
Finally, we briefly describe the three prompt templates that underpin SyntAGM: generation, revision and alignment --- see TA-IV for the complete templates.

\paragraph{Generation}
Builds a few-shot-augmented, grammar-aware prompt that asks the LLM (as a PyOPL expert) to think privately step-by-step and emit only a JSON object with a syntactically valid, literate PyOPL model (.mod) and matching data (.dat). PyOPL syntax/semantics reference and top-$k$ exemplars are injected to guide model development. 

\paragraph{Revision}
Constructs a repair prompt that preserves modelling intent while fixing either (i) syntax/semantic issues (compiler diagnostics) or (ii) misalignment (assessor critique) by introducing minimal edits. It includes the previous attempt, feedback, grammar reference, and few-shots, and again requests only the JSON with full model \& data. 

\paragraph{Alignment}
Elicits a judgement of whether model and data match the problem intent. Returns a JSON object with a boolean ``aligned'' flag and a concise assessment, conditioning on the problem text, grammar reference, and the candidate artefacts produced in the latest attempt. 

\section{Computational study}

\subsection{Baselines}

As remarked in \cite{DBLP:conf/ijcai/XiaoXXGZHFYWSKD25}, many GenMP methods remain closed-source; other methods are interactive, and thus not directly comparable to end-to-end approaches. Moreover, our study aims to produce an AML output; and therefore some baselines such as CAFA \cite{deng2024cafa} and ORLM \cite{Huang2025} --- which are trained to produce models expressed via Python APIs --- are not directly comparable nor applicable.  We therefore limit our study to the following open-source baselines, which have been adapted to produce PyOPL outputs:\footnote{Since the grammar of PyOPL is brand new, it would be unreasonable to expect baselines to produce valid PyOPL models without guidance; therefore, to be fair to all approaches, we always inject the PyOPL grammar in the prompt.} {\bf Standard}, a single-pass zero-shot (instruction-only) strategy; {\bf Chain-of-Thought} (CoT) \cite{10.5555/3600270.3602070}; {\bf Tree-of-Thoughts} (ToT) \cite{10.5555/3666122.3666639}; {\bf Reflexion} \cite{NEURIPS2023_1b44b878}; and {\bf Chain-of-Experts} (CoE) \cite{xiao2024chainofexperts} --- detailed baseline descriptions are provided in TA-VII.

\subsection{Datasets}

We utilise four benchmarks drawn from the unified, cleaned collection of optimisation modelling benchmarks of \cite{DBLP:conf/ijcai/XiaoXXGZHFYWSKD25}: NL4Opt \cite{ramamonjison-etal-2022-augmenting} (214 instances); ComplexOR \cite{xiao2024chainofexperts} (18 instances); ReSocratic \cite{DBLP:conf/iclr/Yang0HGSHFSL025} (351 instances); IndustryOR \cite{Huang2025} (42 instances).
These benchmarks cover different degrees of complexity; and are listed by increasing complexity. Each benchmark is described in detail in TA-VIII. 
Since the cleaned version of ComplexOR and IndustryOR comprise a limited number of instances, we cluster these two benchmarks into a single new benchmark labelled ``ChallengeOR.''
Moreover, we augment this new benchmark with 20 problem instances presenting natural language descriptions of complex
two-stage or multi-stage decision problems under uncertainty, accompanied by scenario-based data and ground-truth solutions. 
This is a class of problems that has not been yet considered in GenMP studies.
These instances have been generated by following a strategy akin to 
ReSocratic \cite{DBLP:conf/iclr/Yang0HGSHFSL025}: first we produced a literate 
stochastic programming model in PyOPL, then we back-translated it 
to a natural language question using a back-translation prompt.
Finally, we manually screened the generated questions. 
The resulting ChallengeOR dataset therefore comprises a total of 80 instances.
Because ChallengeOR is smaller than NL4Opt and ReSocratic and partly synthetic, we primarily use it as a stress test for assessing robustness and cost--quality trade-offs, and we interpret accuracy differences on this benchmark more cautiously. 
As transparency/sensitivity breakdown, in TA-X we report separate results for the 20 new problem instances of decision problems under uncertainty.

\subsection{Model setup, metrics and telemetry}

We use the following LLMs from OpenAI: GPT-5.3-Codex-2026-02-05, GPT-5.4-mini-2026-03-17, and their open-weight gpt-oss-20b (2025-08). 
Unless otherwise stated, we use provider-default decoding/inference parameters to reflect standard deployment conditions; we do not set max\_output\_tokens or custom stop sequences. Results are from single runs unless noted.
Full model and system setup details are provided in TA-IX. For SyntAGM, the maximum number of iterations is set to 5; RAG returns the top $k=3$ results for few-shot prompting --- the exemplar knowledge base is disjoint from all benchmark test instances. 

We adopt the following metrics to compare our approach against baselines. 
The accuracy is computed as the proportion of instances where the observed objective value and the benchmark ground truth are either both ``null'' or they are numerical and close within a given tolerance (relative 1e-6, absolute 1e-9).
Note that although SyntAGM uses an LLM-based judge during inference, its outputs are not used to compute the reported experimental metrics, which are evaluated externally.
We also report the proportion of instances with compilation errors (CE), runtime errors (RE), and wrong answers (WA), where the discrepancy between observed and expected objective values exceeds the given tolerances.
For all approaches we record telemetries in the form of number of iterations per request, latency (in seconds) measured as end-to-end wall-clock time per request, total number of tokens (prompt \& completion) used to answer a request, and associated cost (\$) for pay-per-use models. 

Existing GenMP studies primarily focus on and report accuracy (or related formulation-quality measures) as the gold standard for comparing approaches. 
In practice, however, usability also depends on latency, token consumption, and monetary cost, especially for multi-step pipelines that involve compilation retries, alignment checks, and multi-agent exchanges.
A more nuanced, Pareto-driven, analysis of the cost incurred to achieve a certain level of accuracy is necessary. 
Departing from the approach commonly adopted in the literature, our computational study explores these different dimensions through a cost--quality lens rather than accuracy alone.

\subsection{Results and discussion}

We next analyse the results of our computational study. 

{\bf [gpt-oss:20b]}
Table \ref{tab:nl4opt} reports results on NL4Opt using gpt-oss-20b.
Because NL4Opt comprises 214 cleaned test instances, we treat the aggregate accuracy figures as substantive empirical evidence; additionally,  we include Wilson 95\% confidence intervals (CI) over instances ($n=214$) to quantify uncertainty with respect to the benchmark sample. 
To be fair to all approaches, we included compilation and alignment checks in the loop, so that if at the end of an iteration a model compiles and is judged to be aligned by the LLM, modelling stops. The accuracy-based ranking obtained is broadly aligned with previous studies \cite{xiao2024chainofexperts,DBLP:conf/ijcai/XiaoXXGZHFYWSKD25}. 
In this setting, SyntAGM attains the highest observed accuracy (77.1\%) while also requiring substantially fewer prompt/completion tokens and markedly lower latency than CoE, the strongest competing multi-agent baseline. 
CoE remains competitive in accuracy, but its agentic orchestration induces very large token and time overheads: one needs to wait an average of 10 minutes for model \& data to be produced; SyntAGM requires about 2 minutes, and Reflexion 2 minutes and a half. 
Standard prompting is cheaper but substantially less accurate. 

\begin{table}[htbp]
\centering
\resizebox{0.45\textwidth}{!}{%
\begin{tabular}{|l|c|c|c|c|c|c|}
\hline
\textbf{Method} & \multicolumn{6}{c|}{\textbf{NL4Opt}} \\ \cline{2-7}
& Accur. (95\% CI)& CE \% & RE \% & WA \% & Avg. P/C Tkns & Avg. L (s) \\ \hline
Standard         & 30.3 (24.6--36.8) & 49.0 & 17.2 & 3.27 & 6.92k/2.16k & 74  \\ \hline  
Chain-of-Thought & 46.7 (40.2--53.4) & 1.87 & 46.2 & 5.14 & 6.52k/2.14k & 122 \\ \hline  
Tree-of-Thoughts & 36.0 (29.9--42.6) & 15.4 & 45.7 & 2.80 & 10.6k/5.16k & 224 \\ \hline  
Reflexion        & 59.8 (53.1--66.2) & 1.40 & 33.1 & 5.61 & 11.6k/4.22k & 158 \\ \hline  
Chain-of-Experts & 69.1 (62.7--75.0) & 2.80 & 21.0 & 7.01 & 59.1k/15.7k & 651 \\ \hline  
SyntAGM          & 77.1 (71.0--82.2) & 0.93 & 13.0 & 8.88 & 9.83k/4.16k & 131 \\ \hline  
\end{tabular}
}
\caption{NL4Opt (gpt-oss-20b); Avg. P/C Tkns: average (P)rompt/(C)ompletion tokens; L: latency in seconds}
\label{tab:nl4opt}
\end{table}

A similar pattern emerges in Table \ref{tab:ReSocratic} for ReSocratic, which comprises 351 instances and therefore provides further substantial empirical evidence beyond NL4Opt. SyntAGM again attains the highest observed accuracy (74.9\%), while using far fewer tokens and lower latency than CoE. Reflexion remains competitive but is slower and less accurate. 

\begin{table}[htbp]
\centering
\resizebox{0.45\textwidth}{!}{%
\begin{tabular}{|l|c|c|c|c|c|c|}
\hline
\textbf{Method} & \multicolumn{6}{c|}{\textbf{ReSocratic}} \\ \cline{2-7}
& Accur.\ (95\% CI) & CE \% & RE \% & WA \% & Avg.\ P/C Tkns & Avg.\ L (s) \\ \hline
Standard         & 22.7 (18.6--27.4) & 60.9 & 12.2 & 3.99 & 7.21k/2.78k & 82.7 \\ \hline
Chain-of-Thought & 45.5 (40.4--50.7) & 0.85 & 47.2 & 6.27 & 6.39k/2.39k & 129 \\ \hline
Tree-of-Thoughts & 33.0 (28.3--38.1) & 13.9 & 47.8 & 5.12 & 10.3k/5.13k & 234 \\ \hline
Reflexion        & 61.5 (56.3--66.4) & 3.99 & 28.4 & 5.98 & 12.8k/5.47k & 203 \\ \hline
Chain-of-Experts & 66.1 (61.0--70.9) & 4.84 & 21.9 & 7.12 & 65.6k/16.4k & 651 \\ \hline
SyntAGM          & 74.9 (70.1--79.2) & 5.41 & 11.1 & 8.55 & 12.0k/5.53k & 159 \\ \hline
\end{tabular}
}
\caption{ReSocratic (gpt-oss-20b)}
\label{tab:ReSocratic}
\end{table}

Since, in the gpt-oss-20b experiments on NL4Opt and ReSocratic, CoT and ToT are clearly dominated by stronger baselines in both observed accuracy and efficiency, we exclude these methods from the remainder of the analysis, while retaining Standard as a lightweight reference baseline.

\begin{table}[htbp]
\centering
\resizebox{0.45\textwidth}{!}{%
\begin{tabular}{|l|c|c|c|c|c|c|}
\hline
\textbf{Method} & \multicolumn{6}{c|}{\textbf{ChallengeOR}} \\ \cline{2-7}
& Accur.\ (95\% CI) & CE \% & RE \% & WA \% & Avg.\ P/C Tkns & Avg.\ L (s) \\ \hline
Standard         & 7.50 (3.48--15.4)  & 30.0 & 62.5 & 0.00 & 3.07k/2.20k  & 236  \\ \hline
Reflexion        & 18.7 (11.7--28.7)  & 5.00 & 76.2 & 0.00 & 4.65k/3.11k  & 376  \\ \hline
Chain-of-Experts & 17.5 (10.7--27.3)  & 10.0 & 68.7 & 3.75 & 48.9k/13.6k  & 1089 \\ \hline
SyntAGM          & 30.0 (21.1--40.8)  & 7.50 & 56.2 & 6.25 & 7.98k/7.56k  & 481  \\ \hline
\end{tabular}
}
\caption{ChallengeOR (gpt-oss-20b)}
\label{tab:ChallengeOR-gpt-oss-20b}
\end{table}

Table \ref{tab:ChallengeOR-gpt-oss-20b} reports results on ChallengeOR, which is smaller, harder, and partly synthetic; accordingly, we interpret accuracy rankings more cautiously. 
The benchmark serves as a useful stress test because all methods exhibit a pronounced drop in accuracy relative to the larger public benchmarks. In this more demanding regime, SyntAGM achieves the highest observed accuracy (30.0\%) while continuing to use far fewer tokens and lower latency than CoE. We therefore interpret these results not as definitive accuracy dominance, but as support for SyntAGM's ability to remain competitive under increased difficulty while retaining a better efficiency profile.

{\bf [GPT-5.3-Codex]} 
Table \ref{tab:challengeor_53_codex} evaluates ChallengeOR using GPT-5.3-Codex, a proprietary coding-focused model. 
Standard achieves the lowest observed accuracy, showing that bespoke inference frameworks remain beneficial even in the presence of a coding-oriented backbone. 
Reflexion, CoE, and SyntAGM achieve comparable accuracy.
The main finding in this setting is therefore one of efficiency rather than strict accuracy ranking: SyntAGM matches the best observed accuracy while producing lower cost and latency than other strong baselines.

\begin{table}[htbp]
\centering
\resizebox{0.45\textwidth}{!}{%
\begin{tabular}{|l|c|c|c|c|c|c|}
\hline
\textbf{Method} & \multicolumn{6}{c|}{\textbf{ChallengeOR}} \\ \cline{2-7}
& Accuracy& CE & RE & WA & Avg. Cost & Avg. L (s) \\ \hline
Standard & 52.5\% & 35.0\% & 5.00\% & 7.5\% & \$0.026181 & 13.0 \\ \hline
Reflexion & 77.5\% & 2.5\% & 1.25\% & 18.7\% & \$0.091497 & 51.5 \\ \hline
Chain-of-Experts & 78.7\% & 5.00\% & 3.75\% & 12.5\% & \$0.367764 & 144  \\ \hline
SyntAGM & 78.7\% & 1.25\% & 6.25\% & 13.7\% & \$0.069069 & 35.5 \\ \hline
\end{tabular}
}
\caption{ChallengeOR (GPT-5.3-codex)}
\label{tab:challengeor_53_codex}
\end{table}

Rather than reporting CIs, in this setting we conduct a stability audit: we consider 15 randomly sampled instances (5 from each benchmark) and run 20 replications to assess SyntAGM robustness to decoding stochasticity. Table~\ref{tab:success-prop-15} show high stability, with instability concentrated in two instances with substantially lower success probabilities. Manual inspection reveals a consistent bifurcation between two plausible formulations (e.g., continuous vs. integer domains), attributable to ambiguity in the natural-language problem statement. The remaining failures are sporadic compilation errors.

\begin{table}[h]
\centering
\resizebox{0.45\textwidth}{!}{%
\small
\begin{tabular}{*{1}{|l|}*{5}{c}|*{5}{c}|*{5}{c}|}
\hline
&\multicolumn{5}{c|}{\textbf{NL4OPT}} & \multicolumn{5}{c|}{\textbf{ReSocratic}} & \multicolumn{5}{c|}{\textbf{ChallengeOR}}  \\
Instance \# &1 & 2 & 3 & 4 & 5 & 6 & 7 & 8 & 9 & 10 & 11 & 12 & 13 & 14 & 15 \\
\hline
SyntAGM&0.95 & 0.7 & 1 & 1 & 1 & 0.95 & 1 & 1 & 1 & 0.6 & 1 & 1 & 1 & 0.95 & 0.85 \\
\hline
\end{tabular}
}
\caption{Stability audit: empirical success proportion (GPT-5.3-Codex).}
\label{tab:success-prop-15}
\end{table}

{\bf [GPT-5.4-mini]} Table \ref{tab:challengeor_54_mini} reports ChallengeOR results using GPT-5.4-mini, a smaller proprietary model. 
As expected, the smaller model lowers accuracy across the board. 
Even in this setting, SyntAGM attains the highest observed accuracy (60.0\%) while also keeping observed cost and latency low. 
Given the smaller size and stress-test role of ChallengeOR, we refrain from interpreting this result as definitive evidence of accuracy superiority; rather, it reinforces the broader conclusion that SyntAGM achieves a strong accuracy-efficiency trade-off even with a lighter backbone.

\begin{table}[htbp]
\centering
\resizebox{0.45\textwidth}{!}{%
\begin{tabular}{|l|c|c|c|c|c|c|}
\hline
\textbf{Method} & \multicolumn{6}{c|}{\textbf{ChallengeOR}} \\ \cline{2-7}
& Accuracy& CE & RE & WA & Avg. Cost & Avg. L (s) \\ \hline
Standard & 31.2\% & 47.5\% & 3.75\% & 17.5\% & \$0.014719 & 6.90 \\ \hline
Reflexion & 38.7\% & 35.0\% & 17.5\% & 8.75\% & \$0.082784 & 42.2 \\ \hline
Chain-of-Experts & 51.2\% & 27.5\% & 11.2\% & 10.0\% & \$0.344987 & 141  \\ \hline
SyntAGM & 60.0\% & 18.7\% & 3.75\% & 17.5\% & \$0.077193 & 32.2 \\ \hline
\end{tabular}
}
\caption{ChallengeOR (GPT-5.4-mini)}
\label{tab:challengeor_54_mini}
\end{table}

Taken together, the computational study supports a consistent pattern. On the larger public benchmarks, NL4Opt and ReSocratic, SyntAGM attains the highest observed accuracy while also reducing latency and token usage relative to the strongest multi-agent baseline. On the smaller and more demanding ChallengeOR stress test, accuracy differences among the strongest methods should be interpreted more cautiously; the main result there is that SyntAGM achieves a considerable reduction in effort (i.e. latency and cost/tokens) whilst maintaining competitive observed accuracy. 

\subsection{Ablation study}

We carry out an ablation study to probe the impact of the main SyntAGM components on accuracy and related telemetry. 
The study is performed on ChallengeOR using GPT-5.4-mini, which provides a representative difficult setting.
More specifically, we consider the following components: 
\paragraph{BNF} in-context BNF grammar document. 
\paragraph{RAG} in-context examples via RAG.
\paragraph{Alignment} alignment checks.
\paragraph{Diag. Full/Partial/none} actionable compiler diagnostics. In particular, under ``Diag. (none)'' the compiler simply returns ``Syntax error'' if a compilation fails; under ``Diag. (Partial)''  the compiler returns the line number where the syntax error occurs; under ``Diag. (Full),'' in addition to the line number, the compiler also provide a specific message describing the nature of the issue that led to the syntax error and how the error can be resolved.
\paragraph{Literate} literate-style comments in the generated models and in examples from the RAG knowledge base.

We report two complementary ablation studies. 
First, in TA-XI we report standard leave-one-out ablations from the full baseline system, which indicate how much performance degrades when one component is removed while the others are kept in place. 
Second, in Table \ref{tab:ablation_2} we consider a hard-ablation setting in which all components are removed simultaneously and then reintroduced one at a time. This highlights which signals are sufficient to recover performance from a severely degraded starting point.
Note that we do not carry out an experiment in which we disable PyOPL compiler feedback: the compiler is the ground-truth validator and reward signal for the iteration loop; removing it turns the task from ``produce code that compiles and matches the prompt'' into ``convince an LLM judge,'' which is a different problem.

The study reveals two complementary patterns. 
Standard leave-one-out ablations suggest partial redundancy among BNF, few-shot retrieval, alignment critique, and compiler feedback.
Under a hard-ablation setting, accuracy collapses to 10.0\% and failures are dominated by compilation errors. This is an important result, as it indicates that the underlying LLM is not reliably able to produce valid PyOPL models without external structure and feedback --- and the few ``successes'' are likely obtained by chance due to PyOPL similarity to OPL. Reintroducing individual components from this stripped-down setting yields substantial recoveries: retrieved exemplars via RAG provide the largest single gain (+32.5\%), followed by the in-context BNF grammar reference\footnote{Note that, as one would expect, reintroducing the BNF grammar leads to a performance that is broadly in line with the Standard method.} (+27.5\%) and full actionable diagnostics (+21.2\%). 
It is interesting to note that reintroducing the BNF grammar substantially reduces compilation errors, but this does not always translate to correct models, as witnessed by the high rate of WA. Conversely, reintroducing RAG examples has a less marked effect on compilation errors; this is nevertheless compensated by a reduced rate of WA, leading to an overall superior performance in terms of accuracy.
The comparison between partial and full compiler diagnostics is particularly informative. Reporting only the location of a syntax error has negligible effect relative to the no-guidance setting, whereas full actionable diagnostics produce a marked improvement. This suggests that the main benefit comes not from merely signalling failure, but from providing specific repair information that the model can act upon in subsequent iterations.
Alignment checks provide a smaller but still positive gain (+3.7\%). 

\begin{table}[htbp]
\centering
\resizebox{0.45\textwidth}{!}{%
\begin{tabular}{|l|c|c|c|c|c|c|c|}
\hline
 & \multicolumn{7}{c|}{\textbf{ChallengeOR}} \\ \cline{2-8}
\textbf{Retained}& Accuracy& CE & RE & WA & Avg. Cost & Avg. L (s) & Avg. I\\ \hline
None & 10.0\% & 88.7\% & 0.00\% & 1.25\% & \$0.023154 & 23.5 & 4.62  \\ \hline
BNF & 37.5\% & 27.5\% & 16.2\% & 18.7\% & \$0.033723 & 17.3 & 2.49  \\ \hline
RAG & 42.5\% & 46.2\% & 2.50\% & 8.75\% & \$0.029195 & 23.4 & 2.96  \\ \hline
Alignment & 13.7\% & 83.7\% & 0.00\% & 2.50\% & \$0.021952 & 24.0 & 4.64  \\ \hline
Diag. (Full)& 31.2\% & 50.0\% & 7.50\% & 11.2\% & \$0.019838 & 24.0 & 3.81  \\ \hline
Diag. (Partial)& 11.2\% & 87.5\% & 0.00\% & 1.25\% & \$0.022825 & 24.0 & 4.58  \\ \hline
\hline
Baseline & 60.0\% & 18.7\% & 3.75\% & 17.5\% & \$0.077193 & 32.2 & 3.95  \\ \hline
\end{tabular}
}
\caption{Hard-ablation study on SyntAGM components (GPT-5.4-mini); Avg. I: average iterations}
\label{tab:ablation_2}
\end{table}

\begin{figure*}[htbp]
  \centering
  \includegraphics[width=0.81\textwidth]{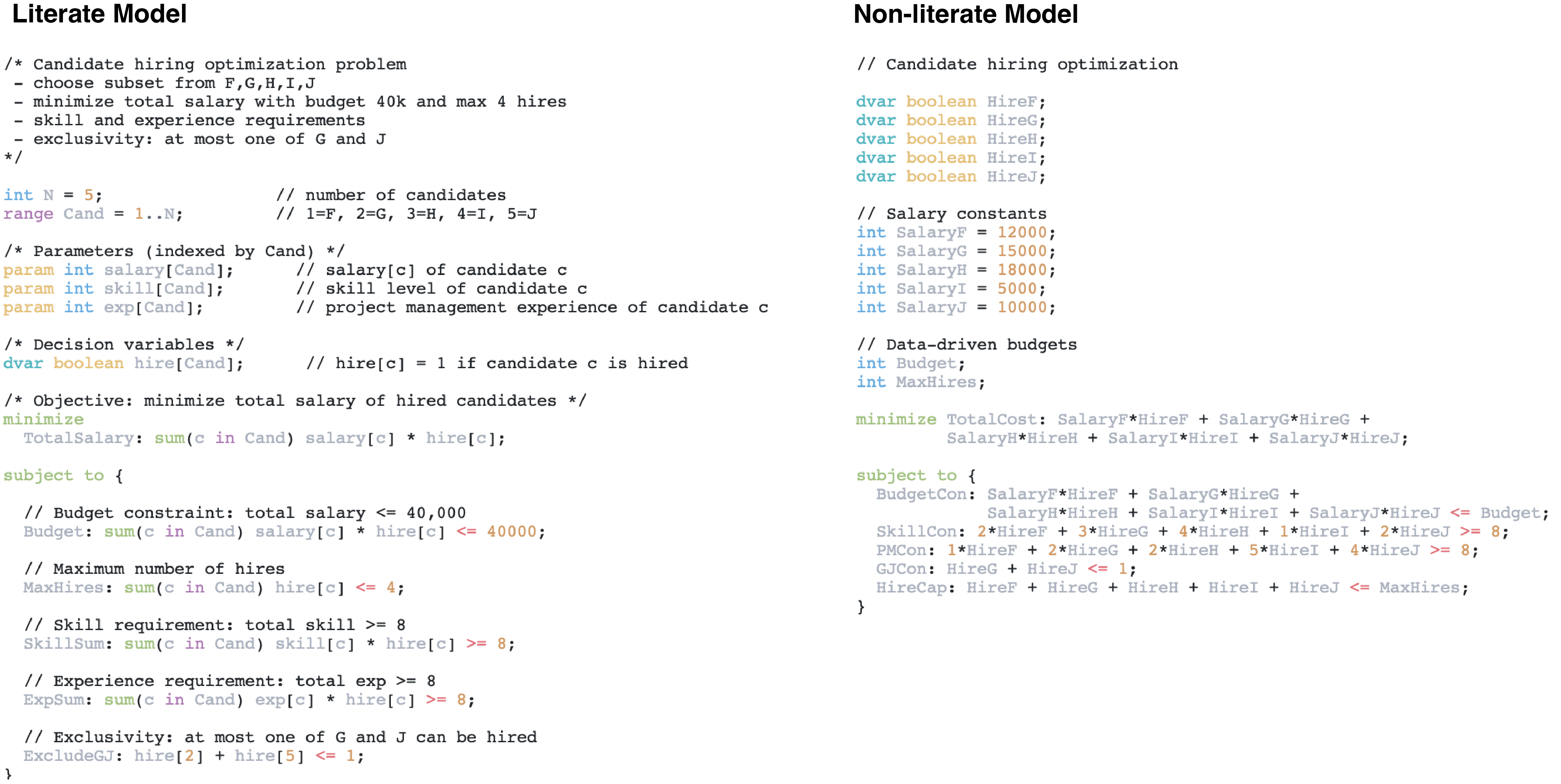}
  \caption{A comparison illustrating the impact of literate-style comments on model readability.}
  \label{fig:literate_ablation}
\end{figure*}

Finally, in our study ablation of literate-style comments in the generated models and of literate-style comments from examples in the RAG knowledge base did not produce differences in accuracy with respect to the baseline. 
However, inspection of models generated (Figure \ref{fig:literate_ablation}) suggests that literate-style comments induce better modelling practices (e.g. appropriate use of indexes, model-data separation) and make models more readable and auditable.

\section{Conclusion and Future Work}

We introduced SyntAGM, a system that pairs a concrete PyOPL syntax/semantics reference document with a compiler-in-the-loop generate--compile--assess--revise workflow. 

Our computational study indicates competitive accuracy and materially lower token, cost, and latency profiles relative to the prompting baselines considered. 
Literate modelling further supports readability and reuse, while telemetry exposes cost--quality trade-offs often overlooked by prior work. 

This study has several limitations.
Our results are established in a controlled PyOPL-based AML synthesis setting and focus on MILP constructs;
our evaluation remains primarily objective-wise and therefore does not fully capture formulation equivalence or human-perceived model quality.

Future works may investigate the application of SyntAGM to other AMLs (e.g. GAMS, AMPL) and to different modelling paradigms (e.g. Constraint Programming), as well as evaluation protocols that better capture formulation correctness. 

\section*{Code availability}

A complete implementation of the system discussed is available at \url{https://gwr3n.github.io/rhetor/}.

\section*{Ethical Statement}

All datasets used contain only publicly available mathematical programming instances.

\section*{Acknowledgments}
This material is based upon works supported by the Science Foundation Ireland under Grant No. 12/RC/2289-P2 which is co-funded under the European Regional Development Fund.

\appendix

\section{MILP modelling patterns}

A range of MILP modelling patterns commonly found in mathematical programs is provided in Table \ref{tab:milp_modelling_techniques}.
\begin{table*}[h!]
\centering
\resizebox{\textwidth}{!}{%
\begin{tabular}{|l|p{6.5cm}|p{5.5cm}|}
\hline
\textbf{Technique} & \textbf{Mathematical Formulation} & \textbf{Description} \\
\hline
Logical NOT & $\text{not } y = 1 - y$ & Binary inversion: if $y = 1$, then $\text{not } y = 0$. \\
\hline
Logical AND & 
$z \leq x$, $z \leq y$, $z \geq x + y - 1$ 
 & $z = 1$ only if both $x = 1$ and $y = 1$. \\
\hline
Logical OR & $z \geq x$, $z \geq y$, $z \leq x + y$ & $z = 1$ if either $x = 1$ or $y = 1$. \\
\hline
Bipartite matching & $\sum_{j} x_{ij} = 1\ \forall i,\quad \sum_{i} x_{ij} = 1\ \forall j$ & Assign each left node to exactly one right node and vice versa (classical assignment). \\
\hline
At-least-one coverage constraints & $\sum_{i} a_{ij} x_i \ge 1\ \forall j$ & Each element $j$ must be covered by at least one selected set. \\
\hline
Pattern/column selection variables & \begin{tabular}[c]{@{}l@{}}
$\min\ \sum_{p\in P} c_p x_p$ \\
$\text{s.t.}\ \ \sum_{p\in P} A_{rp} x_p \ge b_r\ \forall r,\ x_p\in\{0,1\}$
\end{tabular}& Choose composite patterns (e.g., pairings/routes/sets) to cover atomic requirements; equality gives set partitioning. \\
\hline
Exactly-one coverage constraints & $\sum_{k} x_{jk} = 1\ \forall j$ & Pick exactly one option for each item (e.g., assign one crew to each shift). \\
\hline
Resource capacity constraints & $\sum_{j} a_{ij} x_{ij} \le C_i\ \forall i$ & Upper-bound total assigned workload/usage for each resource. \\
\hline
Demand satisfaction equalities & $\sum_{i} x_{ij} = d_j\ \forall j$ & Exactly meet each demand node via inbound flow. \\
\hline
Supply satisfaction equalities & $\sum_{j} x_{ij} = s_i\ \forall i$ & Exactly satisfy each source's supply via outbound flow. \\
\hline
Subtour elimination (SE) &
\begin{tabular}[c]{@{}l@{}}
$1 \le u_i \le N$ \\
$u_i - u_j + N\,x_{ij} \le N-1\quad( i,j \ge 2,\ i \ne j)$
\end{tabular} &
Prevents subtours using ordering variables (Miller–Tucker–Zemlin). \\
\hline
Capacity-based SE &
\begin{tabular}[c]{@{}l@{}}
$\ell_1 = 0$; $\ \ell_j \ge d_j$ \\
$\ell_j \ge \ell_i + d_j - C\,(1 - x_{ij})$
\end{tabular} &
Accumulates load along arcs to eliminate subtours under capacity (VRP style). \\
\hline
Demand coverage inequalities & $\sum_{i} x_{ij} \ge d_j\ \forall j$ & Meet or exceed each demand node via inbound flow (allows over-supply). \\
\hline
Setup Costs/Batch Sizes & $y \cdot M \geq x$ & Setup cost $M$ incurred if $x > 0$; $y$ is binary. Alternatively, $y$ is integer number of batches, each of size $M$ \\
\hline
Conditional Expression & $x \leq b + M(1 - y)$ & Constraint active only if $y = 1$. \\
\hline
Disjunctive Rules & Multiple “if-then” constraints & Enforces one of several mutually exclusive conditions. \\
\hline
Separation via disjunctive big-$M$ & 
\begin{tabular}[c]{@{}l@{}}
$x \ge y + \delta - M z$ \\
$y \ge x + \delta - M (1 - z)$ \\
$z \in \{0,1\}$, $\delta>0$ 
\end{tabular} &
Enforces $|x-y| \ge \delta$ by selecting one of two separated orders; used for adjacency in colouring and minimal separation constraints (e.g., $\delta=1$ for distinct colours). \\
\hline
Precedence Constraint & $x_i + d_i \leq x_j + M(1 - y)$ & If $y = 1$, operation $i$ precedes $j$. \\
\hline
Min-Max Objective & $z \geq d_i - M(1 - x_i) \quad \forall i$ & Minimize the maximum of selected attributes. \\
\hline
Max-Min Objective & $z \leq d_i + M(1 - x_i) \quad \forall i$ & Maximize the minimum of selected attributes. \\
\hline
Inventory Balance & $I_t = I_{t-1} + x_t - d_t$ & Tracks inventory over discrete time horizon. \\
\hline
Initial state constraints & $I_1 = I_0 + x_1 - d_1$ & Initialises the dynamic state from given initial conditions (period 0 to 1). \\
\hline
Inventory with Backlogs & $I_t = I_{t-1} + x_t - d_t + b_t$ & Includes backlog variable $b_t$. \\
\hline
Stock capacity limits & $I_t \leq C$ & Upper bound on the state variable (e.g., storage/warehouse capacity). \\
\hline
Activity Start/End & $y_t = x_t - x_{t-1}$, $z_t = x_t - x_{t+1}$ & Identifies start/end of activity in time horizon. \\
\hline
\end{tabular}
}
\caption{MILP modelling patterns and their mathematical formulations}
\label{tab:milp_modelling_techniques}
\end{table*} 

\section{JSON parsing and validation}
We use a strict JSON output contract at the prompt layer and a relaxed parser at run time, which means we accept plain JSON or fenced JSON. If a fenced block is present, it is preferred; otherwise, we extract the first balanced \{...\} object from the text. Any leading/trailing prose is discarded by this relaxed extraction.
Generation responses must be a single JSON object with keys \texttt{"model"} and \texttt{"data"} (both strings). Alignment responses must be a single JSON object with keys \texttt{"aligned"} (boolean) and \texttt{"assessment"} (string). Prompts specify \texttt{additionalProperties:false} and show an example of what the output should look like; server-side schema enforcement is not used. 

\section{Retrieval-Augmented Generation (RAG)}

Retrieval-augmented generation is realised as retrieval-augmented few-shot prompting: given a user query, the system
performs a full-scan, embedding-based semantic search over all problem-description \texttt{.txt} files under \texttt{pyopl/opl\_models}. 
Documents and query are encoded with SentenceTransformers (\texttt{sentence-transformers/all-MiniLM-L6-v2}) L2-normalised, and scored by cosine similarity implemented as a dot product, \(s(d,q)=\langle d,q\rangle\) (via \texttt{torch.matmul}). 
The top-\(k\) results, where $k=3$,
are post-processed by locating associated \texttt{.mod} and \texttt{.dat} files; each description--model--data triplet is injected verbatim into the prompt under a \texttt{<few\_shot\_examples>} block with explicit instructions to treat exemplars as guidance rather than templates (e.g., avoid copying variable names). The same retrieval scheme is reused across initial synthesis and subsequent syntax/alignment revision prompts.

\section{SyntAGM implementation details}
Algorithm \ref{alg:syntagm} illustrates the implementation of SyntAGM via pseudo-code. The algorithm proceeds through five stages.

\paragraph{Initialisation (lines \ref{alg:gen:load-grammar}--\ref{alg:gen:init-counters})}
Load the PyOPL grammar, retrieve top-$k$ exemplars for few-shot guidance, and assemble the initial generation prompt. Counters for token usage and placeholders for the evolving model/data strings are zeroed.

\paragraph{Iterative synthesis loop (line \ref{alg:gen:loop})}
Each iteration first asks the LLM to produce a candidate PyOPL model \& data (line \ref{alg:gen:invoke-llm}) and parses the JSON response (line \ref{alg:gen:parse-json}). The candidate is compiled with the PyOPL compiler to capture syntax/semantic issues (line \ref{alg:gen:compile}), and artefacts are persisted to the working paths (line \ref{alg:gen:write-io}).

\paragraph{Assessment path (lines \ref{alg:gen:if-clean}--\ref{alg:gen:break})}
On a clean compile, an explicit alignment check is performed via a dedicated assessment prompt (line \ref{alg:gen:assess}). A positive verdict stores the assessor output and exits early (lines \ref{alg:gen:aligned}–\ref{alg:gen:break}); otherwise, an alignment-focused revision prompt is prepared to guide the next attempt (line \ref{alg:gen:align-revise}).

\paragraph{Revision path (line \ref{alg:gen:syntax-revise})}
If compilation fails, a syntax/semantics repair prompt is synthesised that embeds compiler diagnostics, grammar reference, few-shots, and the prior attempt, requesting minimal edits that preserve intent.

\paragraph{Termination and reporting (lines \ref{alg:gen:final-check}--\ref{alg:gen:return})}
If the loop ends with errors, a final assessment is requested to summarise misalignment (line \ref{alg:gen:final-assess}). Usage is aggregated into cost/telemetry (line \ref{alg:gen:cost}) and the routine returns the assessment with optional run statistics (line \ref{alg:gen:return}).

 \begin{algorithm}[h!]
 \caption{SyntAGM}
 \label{alg:syntagm}
 \begin{algorithmic}[1]
 \REQUIRE prompt, model\_path, data\_path, iterations
 \ENSURE assessment (and optionally statistics)
 \STATE Load PyOPL grammar reference (including BNF grammar). \alglinelabel{alg:gen:load-grammar}
 \STATE Retrieve top-$k$ few-shot exemplars from the knowledge base via semantic search. \alglinelabel{alg:gen:retrieval}
 \STATE Build generation prompt using: problem description, grammar reference, and few-shot exemplars.\alglinelabel{alg:gen:build-gen}
 \STATE Initialize usage counters and placeholders for assessment, model\_code, data\_code. \alglinelabel{alg:gen:init-counters}

 \FOR{$t \leftarrow 1$ to iterations} \alglinelabel{alg:gen:loop}
   \STATE Invoke the LLM to synthesise model+data; record token usage. \alglinelabel{alg:gen:invoke-llm}
   \STATE Parse the JSON response to extract model\_code and data\_code. \alglinelabel{alg:gen:parse-json}
   \STATE Compile model+data with the PyOPL compiler; collect any syntax/semantic errors. \alglinelabel{alg:gen:compile}
   \STATE Write model\_code to \textit{model\_path} and data\_code to \textit{data\_path}. \alglinelabel{alg:gen:write-io}

   \IF{no syntax/semantic errors} \alglinelabel{alg:gen:if-clean}
     \STATE Build an alignment-assessment prompt (problem, grammar, model, data) and query the LLM; record token usage. \alglinelabel{alg:gen:assess}
     \IF{assessment indicates aligned} \alglinelabel{alg:gen:aligned}
       \STATE Store assessment; \textbf{break} \alglinelabel{alg:gen:break}
     \ELSE
       \STATE Build an alignment-revision prompt using the assessment, grammar, and few-shots; set it as the next user prompt. \alglinelabel{alg:gen:align-revise}
     \ENDIF
   \ELSE
     \STATE Build a syntax-revision prompt using the compiler errors, grammar, and few-shots; set it as the next user prompt. \alglinelabel{alg:gen:syntax-revise}
   \ENDIF
 \ENDFOR
 \IF{syntax/semantic errors remain} \alglinelabel{alg:gen:final-check}
   \STATE Request a final assessment from the LLM (problem, grammar, latest model+data); record usage.\alglinelabel{alg:gen:final-assess}
 \ENDIF
 \STATE Estimate costs from cumulative usage and package run statistics. \alglinelabel{alg:gen:cost}
 \RETURN assessment (and, if requested, iterations, errors, usage, and cost estimate) \alglinelabel{alg:gen:return}
 \end{algorithmic}
 \end{algorithm}
 
 \section{Worked example}\label{sec:appendix_numerical_example}

We next illustrate the functioning of SyntAGM via a small worked example.
We prompt the system with the problem description in Figure \ref{prompt:alp_1} and Table \ref{prompt:alp_1_parameters}, taken from ComplexOR.

\begin{figure}[h!]
\centering
\begin{tikzpicture}
\node [greybox] (box){%
    \begin{minipage}{0.9\columnwidth}   	
\scriptsize
\texttt{The Aircraft Landing Problem (ALP) is the problem of deciding a landing time on an appropriate runway for each aircraft in a given set of aircraft such that each aircraft lands within a predetermined time window; and separation criteria between the landing of an aircraft, and the landing of all successive aircraft, are respected. We are given the earliest landing time, latest landing time, target landing time, and penalties for landing before or after the target landing time for each aircraft. There is also a separation time that represents the minimum time required between the landing of two aircraft. The objective of the problem is to minimize the total penalties of landing before or after the target time for each aircraft. The problem includes several constraints. The order constraint ensures that the aircrafts land in a specific order. The separation constraint ensures that there is enough separation time between the landing of aircraft. The lower and upper time window constraints ensure that each aircraft lands within its respective earliest and latest time windows.}
    \end{minipage}
};
\node[fancytitle, right=10pt] at (box.north west) {The Aircraft Landing Problem};
\end{tikzpicture}
\caption{Aircraft Landing Problem description}
\label{prompt:alp_1}
\end{figure}

\begin{table*}[h!]
\scriptsize
\centering
\begin{tabular}{ccccccccc}
\hline
 & & & & & & \multicolumn{3}{c}{\textbf{Separation}} \\
\cline{7-9}
\textbf{Aircraft} & \textbf{Earliest} & \textbf{Latest} & \textbf{Target} & \textbf{Penalty Before} & \textbf{Penalty After} &  \textbf{A1} & \textbf{A2} & \textbf{A3} \\
\hline
A1 & 1 & 10 & 4 & 5 & 10 & 0 & 2 & 3 \\
A2 & 3 & 12 & 8 & 10 & 20 & 2 & 0 & 4 \\
A3 & 5 & 15 & 14 & 15 & 30 & 3 & 4 & 0 \\
\hline
\end{tabular}
\caption{Parameters for the Aircraft Landing Problem instance.}
\label{prompt:alp_1_parameters}
\end{table*}

\begin{figure}[htbp]
\centering
\includegraphics[width=\columnwidth]{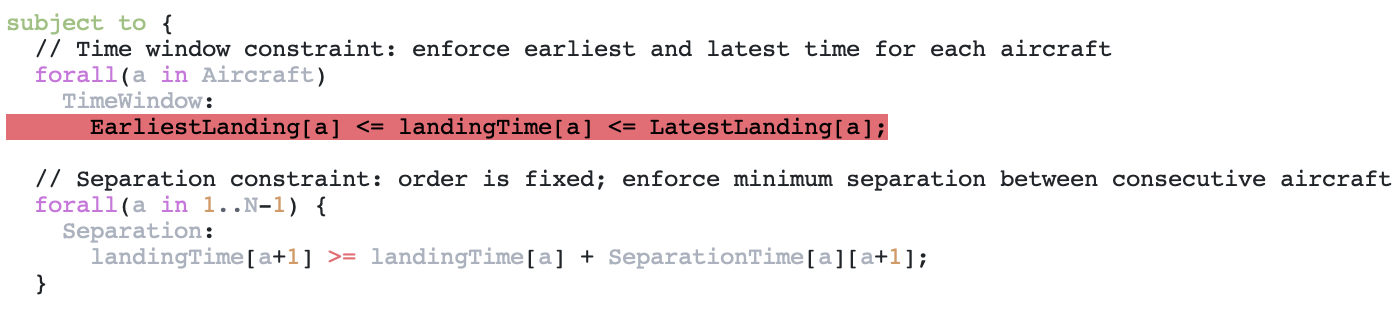}
\caption{Extract from the erroneous PyOPL model for the ALP generated by the system.}
\label{fig:alp_syntax_error}
\end{figure}

The system proceeds by first retrieving $k=3$ few-shot examples: jobshop, crew pairing, and stochastic scheduling.
Then the generation prompt (embedding the BNF grammar) is built and submitted. Model and data file are produced and compiled. At the second iteration round, 
compilation raises the following error: ``Semantic Error (Line 34): Chained comparisons (e.g., $a \leq b \leq c$) are not supported. Split into two constraints: $a \leq b$; $b \leq c$;'' (Figure \ref{fig:alp_syntax_error}) --- note the {\em actionable feedback} provided by the compiler.

Since there are compilation errors, a syntax-revision prompt is built, which incorporates the error message above. This prompt is submitted to initiate the third iteration.
Compilation succeeds and files are assessed for alignment against the original prompt, which is confirmed.

\begin{figure}[htbp]
\centering
\includegraphics[width=\linewidth]{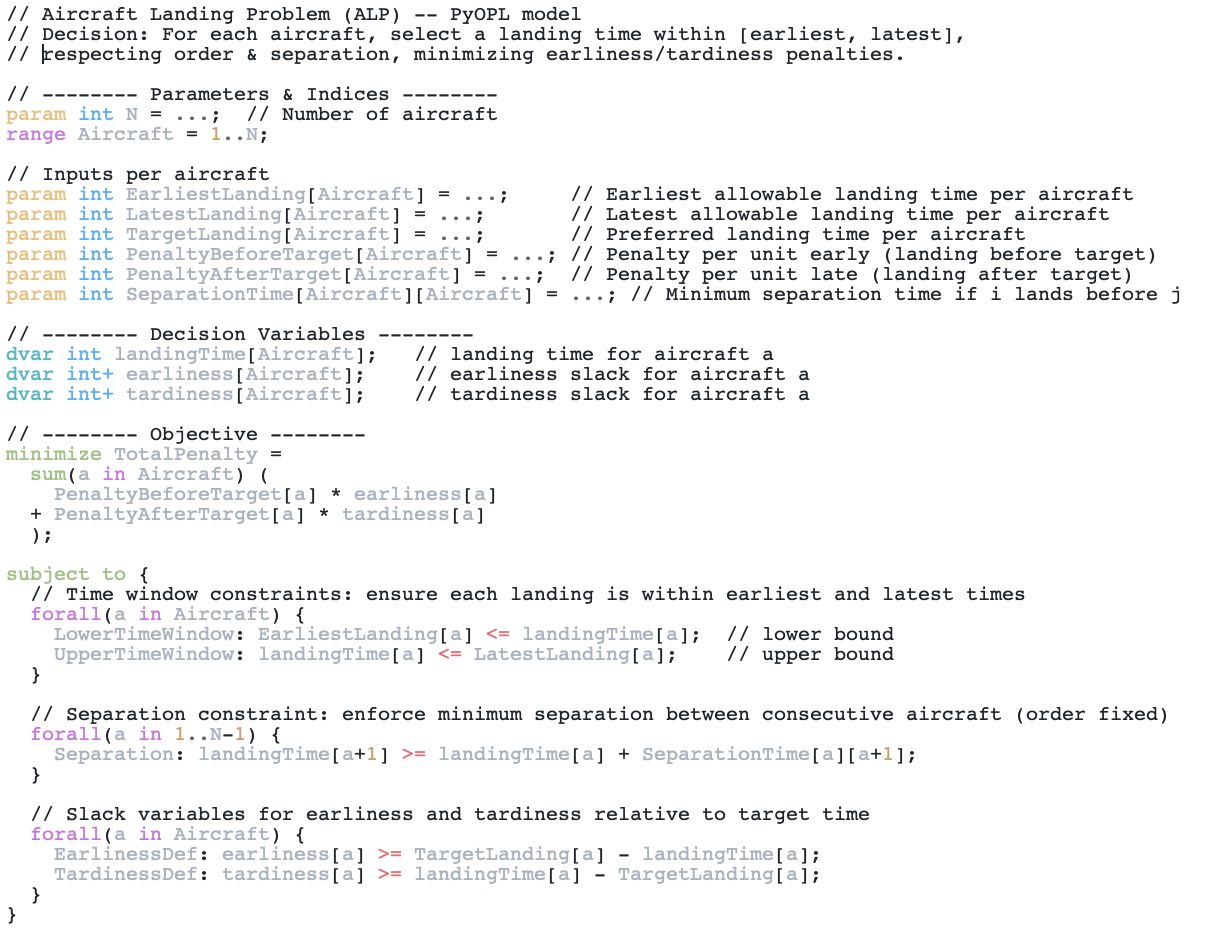}
\caption{PyOPL model for the ALP generated by the system.}
\label{fig:alp}
\end{figure}

After running the model, the optimal solution value (0) matches the ground truth from ComplexOR. The process consumed three iterations, 40638 prompt, and 2179 completion tokens, for a total cost of \$0.081276 on OpenAI GPT-4.1-2025-04-14. A snapshot of the produced PyOPL model with literate style comments is illustrated in Figure \ref{fig:alp}.

A careful analysis reveals that this model is, in fact, not fully aligned with the original problem description. This ought to be expected, as the binary signal produced by the LLM-judge is imperfect and may produce false positives --- this aleatory uncertainty introduced by LLM decoding is thoroughly analysed in our computational study. Moreover, this example also offers a glimpse on a different type of errors that may be observed: although the model produced is not fully aligned with the original problem description, the objective value obtained nevertheless matches the ground truth. In this latter case, because the benchmark instances consist of complex, high-dimensional spaces, in line with other works in the literature we argue that the probability of a random objective-function collision is positive but negligible across benchmark instances, and that the objective-value matching signal serves as a reasonably robust proxy for semantic alignment.

Finally, we also prompt the system a second time to generate an alternative execution path with a different LLM. After generating model and data files, which compile correctly, the system detects a misalignment with the problem description (Figure \ref{prompt:alp_2_misalignment}).

\begin{figure}[h!]
\centering
\begin{tikzpicture}
\node [greybox] (box){%
    \begin{minipage}{0.9\columnwidth}   	
\scriptsize
\texttt{The model and data correctly implement time windows, deviation linking with earliness/lateness, and pairwise separation, and the PyOPL syntax is valid with consistent indices and data dimensions. However, it solves a single‑runway, fixed‑order variant, while the problem description implies assigning each aircraft to an appropriate runway and determining the landing order; these decisions and their variables are absent. To align with the general ALP, introduce runway assignment and sequencing (precedence) binary variables and enforce separation conditionally (e.g., via big‑M) only for consecutive pairs on the same runway. If the intended scope is the fixed‑order single‑runway case, clarify this in the problem statement; in that case the OrderMonotonicity constraints are redundant given nonnegative separation times but harmless. The provided data are consistent with the declared parameters and ranges.}
    \end{minipage}
};
\node[fancytitle, right=10pt] at (box.north west) {Alignment Assessment};
\end{tikzpicture}
\caption{Aircraft Landing Problem --- misalignment with the problem description}
\label{prompt:alp_2_misalignment}
\end{figure}

An alignment-revision prompt is therefore built, which incorporates this assessment. This prompt is submitted to initiate the second iteration.
Compilation succeeds and files are assessed again for alignment against the original prompt, which is confirmed.

After running the model, the optimal solution value (0) matches the ground truth from ComplexOR. The process, which completed in two iterations, consumed 35913 prompt, and 12955 completion tokens, for a total cost of \$0.16696688 on OpenAI GPT-5-2025-08-06.

\section{SyntAGM prompt templates}\label{app:prompt_templates}
In this section, we present detailed descriptions of the three prompt templates the underpin the SyntAGM workflow: generation, revision and alignment.
Text enclosed within double curly braces represent placeholders that are dynamically populated during runtime.

\subsection{Generation} 

{\scriptsize
\begin{lstlisting}[frame=single]
<role>
You are an expert in mathematical optimisation and PyOPL.
</role>

<task>
Think step by step to produce a syntactically valid PyOPL model (.mod) and matching data (.dat) that faithfully implement the problem.
First, reason in a private scratchpad to identify sets, parameters, decision variables, objective, and constraints.
Ensure indices, domains (binary/integer/float), and data are correct and consistent with the problem description.
Choose correct domains (binary/integer/float) from context. Add clear labels and explanatory comments.
Label the objective and each constraint. 
Add concise comments explaining variables, parameters, and constraints, aligned to the problem (literate style).
If any data are missing, create a small, plausible mock instance consistent with the model.
</task>

<grammar_reference>
--- BEGIN PYOPL SYNTAX IMPLEMENTATION ---
{{GRAMMAR_IMPLEMENTATION}}
--- END PYOPL SYNTAX IMPLEMENTATION ---
</grammar_reference>

{{FEW_SHOT_EXAMPLES_SECTION}}

<problem_description>
{{PROMPT}}
</problem_description>

<output_requirements>
- Output ONLY the final JSON with the model and data; do not include your scratchpad in the output.
- Return ONLY a JSON object with keys "model" and "data". Values are single strings; escape quotes and backslashes; encode newlines as \n. No extra keys.
- You MAY wrap the JSON in a ```json fence containing only the JSON.
</output_requirements>

<json_schema>
{ "type": "object", "additionalProperties": false, "required": ["model","data"],
  "properties": { "model":{"type":"string"}, "data":{"type":"string"} } }
</json_schema>

<example_output>
{ "model": "// minimal example\\nfloat a;\\nfloat b;\\ndvar float x;\\nminimize z: a*x;\\nsubject to {\\n  c1: b*x >= 0;\\n}\\n", "data":  "a = 10;\\n b = 5;" }
</example_output>
\end{lstlisting}
}

\subsection{Revision}
{\scriptsize
\begin{lstlisting}[frame=single]
<role>
You are an expert in mathematical optimisation and PyOPL.
</role>

<task>
Revise the model/data to resolve the specified issues while preserving the intended formulation.
Change only what is necessary; keep syntax valid.
Label the objective and each constraint. 
Add concise comments explaining variables, parameters, and constraints, aligned to the problem (literate style).
Use the PyOPL reference strictly for syntax.
</task>

<revision_guidelines>
{{REVISION_GUIDELINE}}
- Make the minimal set of changes necessary to correct syntax/semantic errors.
- Preserve the original modeling structure when possible.
- Ensure the objective, constraints, indices, and variable domains reflect the problem description.
- Keep syntax strictly valid.
- Return complete model and data strings; do not return diffs.
</revision_guidelines>

<grammar_reference>
--- BEGIN PYOPL SYNTAX IMPLEMENTATION ---
{{GRAMMAR_IMPLEMENTATION}}
--- END PYOPL SYNTAX IMPLEMENTATION ---
</grammar_reference>

{{FEW_SHOT_EXAMPLES_SECTION}}

<problem_description>
{{PROMPT}}
</problem_description>

<previous_attempt>
<model>
{{MODEL_CODE}}
</model>

<data>
{{DATA_CODE}}
</data>
</previous_attempt>

<compiler_errors>
{{COMPILER_ERRORS}}
</compiler_errors>

<alignment_assessment>
{{ASSESSMENT}}
</alignment_assessment>

<output_requirements>
- Return ONLY a JSON object with keys "model" and "data". Values are single strings; escape quotes/backslashes; encode newlines as \n. No extra keys.
- You MAY wrap the JSON in a ```json fence containing only the JSON.
</output_requirements>

<json_schema>
{ "type":"object", "additionalProperties": false, "required":["model","data"],
  "properties": { "model":{"type":"string"}, "data":{"type":"string"} } }
</json_schema>

<example_output>
{
  "model": "// minimal example\\nfloat a;\\nfloat b;\\ndvar float x;\\nminimize z: a*x;\\nsubject to {\\n  c1: b*x >= 0;\\n}\\n",
  "data":  "a = 10;\\n b = 5;"
}
</example_output>
\end{lstlisting}
}

If the revision concerns syntax/semantic errors, then \texttt{\{\{REVISION\_GUIDELINE\}\}} is set to ``\texttt{- Fix the listed syntax/semantic errors}.'' Otherwise, if the revision concerns alignment issues, it is set to ``\texttt{- Address the alignment issues noted in the assessment}.''

\subsection{Alignment}

{\scriptsize
\begin{lstlisting}[frame=single]
<role>
You are an expert in mathematical optimization and PyOPL.
</role>

<task>
Judge if model/data fully align with the problem (objective, constraints, variables, indices, and data consistency).
Be specific and critical.
</task>

<grammar_reference>
--- BEGIN PYOPL SYNTAX IMPLEMENTATION ---
{{GRAMMAR_IMPLEMENTATION}}
--- END PYOPL SYNTAX IMPLEMENTATION ---
</grammar_reference>

<inputs>
<problem_description>
{{PROMPT}}
</problem_description>

<model>
{{MODEL_CODE}}
</model>

<data>
{{DATA_CODE}}
</data>
</inputs>

<assessment_focus>
- Objective and constraints reflect the prompt intent.
- Decision variables have correct domains and indices.
- Data is consistent with sets/parameters used by the model.
- Signs, units, and indexing are correct; no missing links.
- Any critical omissions or extraneous constraints.
- Most impactful improvements if misaligned.
</assessment_focus>

<output_requirements>
- Return ONLY a JSON object with keys "aligned" (boolean) and "assessment" (string, 3--6 sentences). No extra keys.
- You MAY wrap the JSON in a ```json fence containing only the JSON.
</output_requirements>

<json_schema>
{ "type":"object", "additionalProperties": false, "required":["aligned","assessment"],
  "properties": { "aligned":{"type":"boolean"}, "assessment":{"type":"string"} } }
</json_schema>
\end{lstlisting}
}

\section{Baselines}

We next briefly discuss the baselines considered in our study.

\paragraph{Standard} The ``standard'' prompting technique implements a single-pass zero-shot (instruction-only) strategy. We set the iteration budget to 1, disable RAG and final assessment, and provide only task instructions plus the PyOPL grammar reference (no exemplars). In particular, we replace the ``task'' in the generation prompt with: ``Produce a syntactically valid PyOPL model (.mod) and matching data (.dat) that faithfully implement the problem. Ensure the model decision variables, objective function, and constraints fully align with the provided problem description. If data are missing, create a small, plausible mock instance consistent with the model.'' 

\paragraph{Chain-of-Thought (CoT)} CoT prompting \cite{10.5555/3600270.3602070} is similar to the former strategy, but with some nuances. To guide the model's thought process, the prompt embeds the sentence ``Think step by step to derive a correct PyOPL model (.mod) and matching data (.dat) for the problem.'' After that, the model is asked to ``First, reason in a private scratchpad to identify sets, parameters, decision variables, objective, and constraints.'' and then ``output ONLY the final JSON with the model and data; do not include your scratchpad in the output.'' This encourages step-by-step reasoning followed by a summarisation stage, whose aim is to aid convergence towards the desired output. Note that this is CoT-style prompting with hidden reasoning, rather than classic CoT prompting.

\paragraph{Tree-of-Thoughts (ToT)} ToT \cite{10.5555/3666122.3666639} frames reasoning as a tree search over intermediate thoughts. This strategy expands multiple candidate branches, scores them (i.e., via compilation or alignment checks), and beam-selects the most promising one to iteratively refine the solution. Our implementation instantiates ToT as a breadth-bounded search. At each level, the search expands $k=3$ candidates from the current best solution. Candidates receive scores if they compile successfully and are deemed aligned by an LLM-based alignment check. The top $b=3$ are beam-selected for the next level. We track the best model/data at every level, and stops when compilation succeeds and alignment passes or the maximum number of trials is reached.

\paragraph{Reflexion} 
Reflexion \cite{NEURIPS2023_1b44b878} leverages a generate--evaluate--reflect loop. An iteration generates a model \& data pair, then compiles them with the PyOPL compiler, and finally runs an LLM-based alignment check. If the check fails, the system initiates a reflection step to determine causes and possible fixes, then restarts the loop by taking into account recent reflections. Our implementation stores the last $M=5$ reflections, injects the PyOPL grammar into prompts, and stops when compilation succeeds and alignment passes or the maximum number of trials is reached.

\paragraph{Chain-of-Experts (CoE)} CoE features a panel of experts orchestrated by a Conductor to tackle problems. The Conductor's role is to select experts (e.g., terminology interpreter, modelling, data builder, code reviewer) who will contribute to building a forward chain of thoughts; once the chain is built, a backward reflection is leveraged to amend faulty steps \cite{xiao2024chainofexperts}.  The Conductor chooses the next expert on the basis of the problem description, the PyOPL grammar reference, and accumulated comments. Once candidate model \& data are produced, compilation results and an LLM-based alignment check are fed back to agents, so that they can address the failure. Our implementation stores the last $M=10$ expert comments, executes up to $5$ forward steps per trial, and stops when compilation succeeds and alignment passes or the maximum number of trials is reached.

\section{Datasets}

\paragraph{NL4Opt}

This dataset \cite{ramamonjison-etal-2022-augmenting} assembles natural-language descriptions of optimisation 
problems paired with ground-truth mathematical programs and solutions. The dataset originates from the NL4Opt 
competition in NeurIPS 2022 and comprises 1101 elementary-level linear programming (LP) problems spanning common OR themes 
(e.g., allocation, transportation), of which 289 are test samples for performance evaluation. 
The cleaned collection of test samples comprises 214 instances.

\paragraph{ComplexOR} 

This dataset \cite{xiao2024chainofexperts} was assembled by OR experts; it comprises 37 problems from
various sources, including academic papers, textbooks, and real-world industry scenarios. Problems span
topics commonly encountered in OR modelling, from supply chain optimisation and scheduling problems 
to warehousing logistics. The cleaned collection of test samples comprises 18 instances.

\paragraph{ReSocratic}

ReSocratic \cite{DBLP:conf/iclr/Yang0HGSHFSL025} adopts a reverse Socratic approach that generates optimisation problems from 
the answer back to a question. By relying on this framework, the authors synthesise 605 problem instances comprising
linear and nonlinear problems; we limit our analysis on the 351 linear problem instances from this dataset, as PyOPL does not currently support 
nonlinear programming.

\paragraph{IndustryOR}

This dataset \cite{Huang2025} features 100 real-world industry cases that covers a variety of problem types and features descriptions with or without tabular data, thereby increasing problem complexity. The cleaned collection of test samples comprises 42 instances.

\section{Model and system setup}

We use the following LLMs from OpenAI: 
GPT-4.1-2025-04-14 and GPT-5-2025-08-06 for the numerical example in Section \ref{sec:appendix_numerical_example};
GPT-5.3-Codex-2026-02-05, GPT-5.4-mini-2026-03-17, and their open-weight gpt-oss-20b (2025-08) in our computational study. 
We use the OpenAI Python SDK's Responses API for OpenAI-hosted models, and Ollama for gpt-oss-20b.
By default, we do not set decoding parameters such as temperature, top-$p$, presence/frequency penalties, or a random seed; these parameters are omitted from requests and provider defaults apply.
We request JSON output (no server-side schema).
\texttt{max\_output\_tokens} (OpenAI) and \texttt{num\_predict} (Ollama) are left unset.
Stop sequences are not sent. Each request samples a single output.
We run gpt-oss-20b on an Intel(R) Xeon(R) w5-2465X, 3096 MHz, 16 Core(s), 32 Logical Processor(s) with 128 GB of RAM and NVIDIA RTX 2000 Ada Generation with 16 GB GDDR6. Defaults and cost per-1k token rates were captured on May 1, 2026 with openai 2.21.0 and Python 3.11.13. 

\section{StochasticOR}

Results obtained for StochasticOR when using gpt-oss-20b, GPT-5.3-Codex, and GPT-5.4-mini are shown in Tables \ref{tab:stochasticor_53_codex}-\ref{tab:stochasticor_54_mini}.
StochasticOR is a set of 20 problem instances, embedded in ChallengeOR, presenting natural language descriptions of complex two-stage
or multi-stage decision problems under uncertainty

\begin{table}[htbp]
\centering
\resizebox{0.45\textwidth}{!}{%
\begin{tabular}{|l|c|c|c|c|c|c|}
\hline
\textbf{Method} & \multicolumn{6}{c|}{\textbf{StochasticOR}} \\ \cline{2-7}
& Accuracy \% & CE \% & RE \% & WA \% & Avg.\ P/C Tkns & Avg.\ L (s) \\ \hline
Standard         & 0.00 & 10.0 & 90.0 & 0.00 & 0.81k/0.98k  & 379  \\ \hline
Reflexion        & 0.00 & 0.00 & 100 & 0.00 & n/a  & 308  \\ \hline
Chain-of-Experts & 5.0 & 10.0 & 85.0 & 0.00 & 27.0k/11.3k  & 1293 \\ \hline
SyntAGM          & 5.0 & 5.00 & 85.0 & 5.00 & 4.71k/6.36k  & 648  \\ \hline
\end{tabular}
}
\caption{ChallengeOR (gpt-oss-20b)}
\label{tab:ChallengeOR-gpt-oss-20b}
\end{table}

\begin{table}[htbp]
\centering
\resizebox{0.45\textwidth}{!}{%
\begin{tabular}{|l|c|c|c|c|c|c|}
\hline
\textbf{Method} & \multicolumn{6}{c|}{\textbf{StochasticOR}} \\ \cline{2-7}
& Accuracy& CE & RE & WA & Avg. Cost & Avg. L (s) \\ \hline
Standard & 30.0\% & 55.0\% & 15.0\% & 0.00\% & \$0.030857 & 17.7 \\ \hline
Reflexion & 70.0\% & 10.0\% & 0.00\% & 20.0\% & \$0.165232 & 107 \\ \hline
Chain-of-Experts & 85.0\% & 0.00\% & 10.0\% & 5.00\% & \$0.582785 & 246  \\ \hline
SyntAGM & 80.0\% & 5.00\% & 10.0\% & 5.00\% & \$0.113639 & 62.6 \\ \hline
\end{tabular}
}
\caption{StochasticOR (GPT-5.3-codex)}
\label{tab:stochasticor_53_codex}
\end{table}

\begin{table}[htbp]
\centering
\resizebox{0.45\textwidth}{!}{%
\begin{tabular}{|l|c|c|c|c|c|c|}
\hline
\textbf{Method} & \multicolumn{6}{c|}{\textbf{StochasticOR}} \\ \cline{2-7}
& Accuracy& CE & RE & WA & Avg. Cost & Avg. L (s) \\ \hline
Standard & 14.2\% & 66.6\% & 0.00\% & 19.0\% & \$0.017084 & 9.36 \\ \hline
Reflexion & 30.0\% & 50.0\% & 15.0\% & 5.00\% & \$0.098832 & 59.2 \\ \hline
Chain-of-Experts & 40.0\% & 55.0\% & 5.00\% & 0.00\% & \$0.488354 & 210  \\ \hline
SyntAGM & 50.0\% & 40.0\% & 0.00\% & 10.0\% & \$0.103000 & 47.8 \\ \hline
\end{tabular}
}
\caption{StochasticOR (GPT-5.4-mini)}
\label{tab:stochasticor_54_mini}
\end{table}

\section{Ablation study}

Results of our leave-one-out ablation study are illustrated in Table \ref{tab:ablation_1}.

\begin{table}[htbp]
\centering
\resizebox{0.45\textwidth}{!}{%
\begin{tabular}{|l|c|c|c|c|c|c|c|}
\hline
 & \multicolumn{7}{c|}{\textbf{ChallengeOR}} \\ \cline{2-8}
\textbf{Removed}& Accuracy& CE rate& RE rate& WA & Avg. Cost & Avg. L (s) & Avg. I\\ \hline
BNF & 60.0\% & 12.5\% & 6.25\% & 21.2\% & \$0.039884 & 36.4 & 3.36  \\ \hline
RAG & 40.0\% & 30.0\% & 20.0\% & 8.00\% & \$0.055800 & 30.1 & 3.90  \\ \hline
Alignment & 58.7\% & 10.0\% & 10.0\% & 21.25\% & \$0.031938 & 14.3 & 1.75  \\ \hline
Diag. (Partial)& 53.7\% & 32.5\% & 1.25\% & 12.5\% & \$0.074340 & 37.7 & 4.15  \\ \hline
Diag. (none)& 46.2\% & 37.5\% & 5.00\% & 11.2\% & \$0.077084 & 37.7 & 4.11  \\ \hline
\hline
Baseline & 60.0\% & 18.7\% & 3.75\% & 17.5\% & \$0.077193 & 32.2 & 3.95  \\ \hline
\end{tabular}
}
\caption{Leave-one-out ablation study on SyntAGM components (GPT-5.4-mini)}
\label{tab:ablation_1}
\end{table}

\bibliographystyle{named}
\bibliography{gen_mp}

\end{document}

%% file: patterns_table.tex
\begin{table*}[h!]
\centering
\scriptsize
\setlength{\tabcolsep}{5pt}
\renewcommand{\arraystretch}{1.2}
\resizebox{\textwidth}{!}{%
\begin{tabular}{l|*{22}{c}}
                       & \rotatebox{90}{assignment} & \rotatebox{90}{crew\_pairing} & \rotatebox{90}{crew\_scheduling} & \rotatebox{90}{graph\_coloring} & \rotatebox{90}{inventory\_routing} & \rotatebox{90}{jobshop} & \rotatebox{90}{knapsack} & \rotatebox{90}{knapsackp} & \rotatebox{90}{lot\_sizing} & \rotatebox{90}{plant\_location} & \rotatebox{90}{production} & \rotatebox{90}{set\_covering} & \rotatebox{90}{set\_partitioning} & \rotatebox{90}{transportation} & \rotatebox{90}{tsp} & \rotatebox{90}{vehicle\_routing} & \rotatebox{90}{warehouse\_location} & \rotatebox{90}{workforce\_planning} & \rotatebox{90}{p\_dispersion} & \rotatebox{90}{on\_off\_outsourcing} & \rotatebox{90}{stochastic\_production} & \rotatebox{90}{stochastic\_scheduling} \\
\hline
Logical NOT ($1{-}y$) &  &  &  & \cmark &  & \cmark &  &  &  &  &  &  &  &  &  & \cmark &  &  & \cmark & \cmark &  &  \\
Logical AND ($z{=}x\,y$) &  &  &  &  &  &  &  &  &  &  &  &  &  &  &  &  &  &  & \cmark &  &  &  \\
Logical OR &  &  &  &  &  &  &  &  &  &  &  &  &  &  &  &  &  &  &  & \cmark &  &  \\
Bipartite matching & \cmark &  &  &  &  &  &  &  &  &  &  &  &  &  &  &  &  &  &  &  &  &  \\
At-least-one coverage constraints &  & \cmark &  &  &  &  &  &  &  &  &  & \cmark &  &  &  &  &  &  &  &  &  &  \\
Pattern/column selection variables &  & \cmark &  &  &  &  &  &  &  &  &  & \cmark & \cmark &  &  &  &  &  &  &  &  &  \\
Exactly-one coverage constraints & \cmark &  & \cmark &  &  &  &  &  &  &  &  &  & \cmark &  & \cmark & \cmark &  &  &  &  &  &  \\
Resource capacity constraints &  &  & \cmark &  &  & \cmark & \cmark & \cmark &  & \cmark & \cmark &  &  &  &  & \cmark & \cmark & \cmark &  & \cmark &  & \cmark \\
Demand satisfaction equalities &  &  &  &  &  &  &  &  &  & \cmark &  &  &  & \cmark &  &  & \cmark &  &  &  &  &  \\
Demand coverage inequalities &  & \cmark &  &  &  &  &  &  &  &  & \cmark & \cmark &  &  &  &  &  & \cmark &  & \cmark & \cmark & \cmark \\
Supply satisfaction equalities &  &  &  &  &  &  &  &  &  &  &  &  &  & \cmark &  &  &  &  &  &  &  &  \\
Subtour elimination (SE) &  &  &  &  &  &  &  &  &  &  &  &  &  &  & \cmark &  &  &  &  &  &  &  \\
Capacity-based SE &  &  &  &  &  &  &  &  &  &  &  &  &  &  &  & \cmark &  &  &  &  &  &  \\
Setup Costs/Batch Sizes &  &  &  &  &  &  &  &  & \cmark & \cmark &  &  &  &  &  &  & \cmark &  &  & \cmark &  &  \\
Conditional Expression (big-$M$) &  &  &  & \cmark &  & \cmark &  &  & \cmark & \cmark &  &  &  &  & \cmark & \cmark & \cmark &  & \cmark & \cmark &  &  \\
Disjunctive Rules &  &  &  & \cmark &  & \cmark &  &  &  &  &  &  &  &  &  &  &  &  &  &  &  &  \\
Separation via disjunctive big-$M$ &  &  &  &  &  &  &  &  &  &  &  &  &  &  &  &  &  &  &  &  &  &  \\
Precedence Constraint &  &  &  &  &  & \cmark &  &  &  &  &  &  &  &  &  &  &  &  &  &  &  &  \\
Min-Max Objective &  &  &  & \cmark &  & \cmark &  &  &  &  &  &  &  &  &  &  &  &  &  &  &  &  \\
Max-Min Objective &  &  &  &  &  &  &  &  &  &  &  &  &  &  &  &  &  &  & \cmark &  &  &  \\
Inventory Balance &  &  &  &  & \cmark &  &  &  & \cmark &  &  &  &  &  &  &  &  & \cmark &  &  &  & \cmark \\
Initial state constraints &  &  &  &  & \cmark &  &  &  & \cmark &  &  &  &  &  &  &  &  & \cmark &  &  &  & \cmark \\
Inventory with Backlogs &  &  &  &  &  &  &  &  &  &  &  &  &  &  &  &  &  &  &  &  &  & \cmark \\
Stock capacity limits &  &  &  &  & \cmark &  &  &  &  &  &  &  &  &  &  &  &  &  &  &  &  &  \\
Activity Start/End &  &  &  &  &  &  &  &  &  &  &  &  &  &  &  &  &  &  &  & \cmark &  &  \\
\hline
\end{tabular}%
}
\caption{Which modelling patterns are used in each PyOPL problem (\cmark indicates presence).}
\label{tab:patterns_by_problem}
\end{table*}